\documentstyle[12pt,prc,aps]{revtex}
\begin{document}
\begin{titlepage}

\title{\bf Does the Sun Appear Brighter at Night in Neutrinos?}

\author{J.N. Bahcall and P.I. Krastev}
\address{School of Natural Sciences, Institute for Advanced
Study\\
Princeton, NJ 08540\\}
\maketitle
\input psfig

\begin{abstract}

We calculate accurately the
number of solar neutrino events expected 
as a function of solar zenith
angle, with and without 
neutrino oscillations, for detectors at
the locations of Super-Kamiokande, SNO, and the Gran Sasso National
Laboratory.  Using different earth models to
estimate  geophysical uncertainties, and different solar models
to estimate solar uncertainties,
we evaluate distortions predicted by the MSW effect  in the
 zenith angle distributions of solar neutrino events.  The distortions are
caused by oscillations and by $\nu-e$ interactions in the earth that 
regenerate $\nu_e$ from $\nu_\mu$ or $\nu_\tau$.
We show that the first two 
moments of the zenith-angle distribution are more 
sensitive to the small mixing angle MSW solution than
the conventionally studied  day-night
asymmetry.  We present iso-sigma contours that illustrate the potential
of Super-Kamiokande, SNO, BOREXINO, ICARUS and HERON/HELLAZ for
detecting 
the earth regeneration effect at their actual locations 
(and at the equator).
MSW solutions
favored by the four pioneering solar neutrino experiments
predict  characteristic distortions for Super-Kamiokande,  SNO,
BOREXINO, and ICARUS
that range from being
unmeasurably small to  
 $> 5\sigma$ (stat)
after only a few years of observations.
\end{abstract}
\end{titlepage}
\newpage

\section{Introduction}
\label{intro}
Four operating solar neutrino experiments
(Chlorine\cite{homestake}, Kamiokande\cite{kamioka},
GALLEX\cite{GALLEX}) and SAGE\cite{SAGE}
have detected neutrinos from nuclear fusion in the interior of the sun
with approximately the numbers and energies expected from standard solar
models\cite{snp,BP95}.
Moreover, sound speeds calculated from the standard solar models agree
with the helioseismologically determined sound speeds to a rms
accuracy of better than $0.2$\% throughout essentially the entire
sun\cite{helio}. 

Nevertheless, quantitative discrepancies have persisted for almost
three decades between the predictions of the standard solar models and
the observations of solar neutrino experiments\cite{thirty,3SNP,hegger}.
Several suggested modifications of neutrino properties provide
excellent fits to the existing solar neutrino data\cite{howell}.

Are there potential ``smoking gun'' indications of new physics?  Yes,
the most popular neutrino physics solution,   the 
Mikheyev-Smirnov-Wolfenstein (MSW) effect
\cite{msw}, predicts several characteristic and unique phenomena.
The MSW effect explains solar neutrino observations as the result of
conversions in the solar interior of  $\nu_e$ 
produced in nuclear reactions
to the more difficult to detect $\nu_\mu$ or $\nu_\tau$.

Potentially decisive signatures of new physics that 
are suggested by the MSW effect 
include observing that the sun is brighter in
neutrinos at night (the `earth regeneration effect') 
\cite{earthreg,BaltzW,BaltzW2},
detecting distortions in the 
incident solar neutrino energy spectrum\cite{elspectJ},
and observing that the flux of all types of neutrinos exceeds the flux
of just electron type neutrinos\cite{chen,snoJE}.
A demonstration that any  of these 
 phenomena exists would provide evidence for physics beyond the
minimal standard electroweak model.

The regeneration effect is an especially powerful diagnostic of new
physics since no
difference is predicted between the counting rates 
observed during
the day and at night 
(or, more generally, any dependence of the counting rate on the solar
zeith angle) by such popular alternatives
to the MSW effect  as vacuum oscillations~\cite{pontecorvo}, 
magnetic moment
transitions~\cite{okun}, or violations of the equivalence
principle~\cite{gasperini}. 

In this paper, we investigate the sensitivity of 
new solar neutrino experiments,
Super-Kamiokande \cite{superK},
SNO \cite{sno}, ICARUS\cite{icarus}, BOREXINO \cite{borexino}, HERON
\cite{heron} and HELLAZ \cite{hellaz}, to 
the earth regeneration effect.
The MSW effect predicts that, for certain values of the neutrino
masses and mixing angles,   $\nu-e$
interactions in the earth (at night) may convert  $\nu_\mu$ or $\nu_\tau$
from the sun back into the more easily detectable $\nu_e$.

An accurate evaluation of the systematic significance of experimental
results will require the detailed Monte Carlo simulations that will be
carried out by the experimental collaborations;   the collaborations 
 will determine
the best estimates and uncertainties for all  the quantities that
affect the experimental result. These results will then be analyzed
using computer codes that include the experimental details and
which make use of optimal statistical techniques such as maximum
likelihood analysis. In the absence of detailed Monte Carlo
simulations of the experimental characteristics (yet to be
determined), we estimate in this paper the statistical significance of
expected results by comparing the predictions of 
various MSW scenarios with respect to
the no-oscillation scenario including only statistical errors and
analyzing the results with a $\chi^2$ statistic.

The reader who wants to see the approximate power of the new experiments
can turn
immediately to Fig.~\ref{isosigma}, 
which shows the significance level (statistical errors only) with
which new solar experiments could detect the regeneration effect.
Experience with the operating experiments over a period of years may be 
necessary to determine the size of the 
systematic errors.

This paper is organized as follows.
In Sec.~\ref{ERE} we summarize the general features of $\nu_e$
regeneration in the earth.
We  then describe in Sec.~\ref{profiles}
the different models of the earth
used for estimating the uncertainties in the numerical calculations due
to uncertainties in the density profile  and  the
chemical composition of the earth's interior.
In the detailed calculations that follow, we use the differences
between the results obtained
from the different models of the earth to determine the geophysical
uncertainties in the MSW predictions.

After these preliminary considerations, we 
determine in Sec.~\ref{rates} the regions in the 
mixing angle and mass difference plane,
\hbox{$\sin^22\theta$ - $\Delta m^2$},   that are 
 allowed by the latest solar neutrino
data, taking into account the influence of earth regeneration on the
predicted counting rates in the chlorine, Kamiokande, GALLEX, and SAGE
experiments.  In addition to the familiar large mixing angle and small
mixing angle MSW solutions, we find an additional 
 LOW solution (large mixing angle,
smaller neutrino mass difference) that has an acceptable confidence
level  only when earth
regeneration is included in the calculations.

We then investigate the sensitivity of next-generation solar neutrino
experiments to the earth regeneration effect.
We begin by defining and calculating 
the zenith-angle  exposure function in 
Sec.~\ref{exposure}.  This function depends only on the location of
a neutrino detector on the surface of the earth; it is independent of the
characteristics of the detector. 
We also calculate in Sec.~\ref{exposure} the distorted zenith angle
distribution that takes account of regeneration in the earth.

The results for MSW regeneration given previously in the literature
involve making  approximations 
either in the model description of the earth or in the calculation of
the average survival probability after regeneration, or both.
Instead, we integrate numerically the differential equations
describing the evolution of the neutrino states in traversing 
an accurate model of the earth, thereby
avoiding the necessity of arguing that an 
approximation scheme is sufficiently accurate.  
In several tables in this paper, we present numerical results to a
precision of $0.01$\%, an accuracy  much higher 
than can be measured experimentally.  These precise numerical
predictions are given in order to illustrate the small effect
on measurable quantities of 
some of the systematic differences.

We introduce in Sec.~\ref{moments} the first two
moments of the zenith-angle distribution of neutrino events and
calculate the dependence of the moments on neutrino parameters 
for   the new solar neutrino
experiments: Super-Kamiokande, SNO,
ICARUS, BOREXINO and HERON/HELLAZ.
For comparison,  
we calculate accurately in Sec.~\ref{ND} the conventional 
day-night asymmetry; we present values of the 
day-night asymmetry   for the new solar neutrino detectors mentioned
above.

Which characterization is more sensitive to new physics: the moments
of the zenith-angle distribution or the day-night asymmetry?  We
show in Sec.~\ref{besttest} that the moments are more sensitive to
the small mixing angle MSW solution and the day-night asymmetry is
more sensitive to the large mixing angle solution.
Although less statistically powerful 
than the full Monte Carlo simulations that
will be carried out by the experimental collaborations,
the analyses using, e. .g, moments or the day-night asymmetry can be
carried out quickly by theoreticians interested in determining whether
new particle physics scenarios can be tested by the experiments or
whether they are already inconsistent with data that have been published.
For completeness, we present in Sec.~\ref{spectrum} the moments of the
electron recoil energy spectrum for Super-Kamiokande and SNO that were
computed including the earth regeneration effect.

We then  show in Sec.~\ref{modelsensitivity} that the MSW predictions
for regeneration in the earth 
depend only slightly on the adopted  density profile of the earth, 
the  chemical
composition of the earth,  and the details of the solar model.

Following the suggestion of Gelb, Kwong, and Rosen~\cite{GKR},
we calculate  in Sec.~\ref{equator} the increase in the sensitivity
to the regeneration effect that could be achieved by building
detectors at the equator.
We discuss and summarize our main results in Sec.~\ref{discussion}.

\section{The Earth Regeneration Effect}
\label{ERE}

We work in a two-neutrino mixing scheme involving $\nu_e$ (produced in
the sun) and either $\nu_\mu$ or $\nu_\tau$ (produced by
oscillations). Soon after Mikheyev and Smirnov suggested
 the MSW effect \cite{msw} as a
possible solution of the solar neutrino problems, several authors
pointed out \cite{earthreg} that day-night variations of the event rates in
solar neutrino detectors could provide 
spectacular confirmation of the MSW effect and thus of new
physics. 

The MSW solution of the solar neutrino problems requires that  
electron neutrinos produced in nuclear reactions in the center of
the sun are converted to muon or tau neutrinos by interactions with
solar electrons on their way from the interior of the sun 
to the detector on earth.  
The conversion in the sun is primarily a resonance phenomenon, which
occurs at a specific density that corresponds to a definite neutrino
energy (for a specified neutrino mass difference).

During day-time, the higher-energy neutrinos
arriving at earth are mostly $\nu_\mu$ (or $\nu_\tau$) with some
admixture of $\nu_e$. At night-time,  neutrinos must pass
through the earth in order to reach the detector.
As a result of traversing the earth,    the 
fraction of the more easily 
detected $\nu_e$ increases
because of the conversion of $\nu_\mu$ (or $\nu_\tau$) to 
$\nu_e$  by neutrino oscillations.  For the small mixing angle MSW
solution, 
interactions with electrons in the earth increase the effective 
mixing angle and enhance the conversion process.  For the large mixing
angle MSW solution, the conversion of $\nu_\mu$ (or $\nu_\tau$) to 
$\nu_e$ occurs by oscillations that are only slightly enhanced over
vacuum mixing.
This process of increasing in the earth 
the fraction of the neutrinos that are
$\nu_e$ is   called the ``regeneration effect'' and has  the
opposite effect to 
 the conversion of $\nu_e$ to $\nu_\mu$ (or $\nu_\tau$) in
the sun. 

Because of the change of neutrino type  in the earth, the
MSW mechanism predicts that solar neutrino detectors 
should generally measure
higher event rates at night than during day-time.

Figure \ref{geometry} illustrates a schematic
view of a solar neutrino detector at the geographic latitude, $\phi$. 
 Since the earth is spherically symmetric to 
 $O (10^{-2.5})$, it is sufficient to consider the
cross-section slice shown in the figure.\footnote{The polar-equatorial 
asymmetry of the earth changes the predicted MSW regeneration effects
by less than or of order $0.3$\%, which is not observable with
existing or planned detectors.}
Two lines determine the geometry: 
one line defines the zenith direction, and 
the other line is the trajectory of
the neutrino. The zenith angle $\alpha$ ($0^0 <
\alpha < 180^0$) between these two lines specifies the
neutrino trajectory in the earth. The survival probability depends on
the neutrino oscillation parameters, $\Delta m^2$ and $\sin^22\theta$,
on the neutrino energy, $E$, 
and on the path (i.e., $\alpha$) the neutrino travels through the earth. Since 
$\alpha$
 changes due to the apparent motion of the sun, the
neutrino survival probability should change with time as well,
resulting in an asymmetric distortion
of the angular distribution of events.

Real-time detectors, which record the times 
 at which neutrinos interact within the detector, are best suited for
studying the earth regeneration effect. In radiochemical experiments
the time of detection is poorly known, since a typical run usually lasts
between several weeks (gallium) and several months
(chlorine). \footnote{The possibility of making
extractions twice a day and counting separately day-time and
night-time samples in the chlorine and iodine experiments has
been considered carefully. \cite{radiodn}.}  

 Kamiokande, a
real-time neutrino electron scattering detector, did not see any
signal for the earth regeneration effect. 
The Kamiokande collaboration used this non-observation
 to rule out an important region
in parameter space for which the predicted day-night asymmetry, or
zenith-angle dependence, is large\cite{kamdn}. However, the
sensitivity of the Kamiokande detector was insufficient to probe the
full $\Delta m^2$ - $\sin^22\theta$ parameter space for which there might
be an appreciable day-night effect, measurable by Super-Kamiokande or SNO.

There are several calculations in the 
literature \cite{BaltzW3,hata,minn,lisimon,liu} of the expected 
magnitude of the regeneration
effect in future experiments. Different groups
of researchers have used different models of the earth in their
calculations. No quantitative estimate has been made previously of the
sensitivity of the measurable 
quantities to the adopted density profile of the earth;  each 
group has typically presented results using a specific density
profile, often not the best available profile. 
In the subsequent sections, we describe direct numerical
computations of the earth regeneration effect for six different models
of the earth. Thus we quantify the dependence of the calculated
characteristics of the earth regeneration effect on the model of the
earth and exhibit the corresponding uncertainties which turn out to be
rather small. The density profiles in the six models of the earth are
described in Sec.~\ref{profiles}.

\section{Earth Models}
\label{profiles}

The MSW effect in the earth depends upon the electron number density
as a function of radius.  In this section, we determine best-estimates
and a range of uncertainties for the total mass density and for the 
chemical composition.  We use the best available earth models and chemical
composition for most of the calculations performed in this paper, but
we also carry out calculations for five older mass models
in order to determine the uncertainties in the
predicted MSW effects that arise from uncertainties in the model of
the earth's density profile.
We use a best-estimate 
chemical composition for the core that is inferred from the
seismological measurements.  To test  the
sensitivity of the MSW predictions to the assumed 
chemical composition of the earth, we make extreme assumptions that
maximize or minimize the average charge to mass ratio and carry out 
calculations also for these extreme cases.

\subsection{Density Profiles}
\label{densities}

The density distribution inside the earth is known
with a precision of a few percent\cite{zharkov}. A large
set of seismic measurements has been used to
obtain the most 
accurate model, PREM\cite{PREM} (the Preliminary
Reference Earth Model), for the earth's density distribution.  
We will use the PREM model for all of our best-estimate calculations.
This model has also been used by Lisi and Montanino~\cite{lisimon} as
the basis of their recent analytic study of earth regeneration. 
 Other models are described in Refs.~\cite{bullen} and
\cite{stacey}. To determine the sensitivity to the assumed density
profile, we  have performed calculations with a representative
set of six different
earth models, all spherically symmetric and with the same
radius, $R_\oplus = 6371$ km.  

Figure \ref{density} shows the density
profiles of the six earth models.  The density distributions in
these models are divided into five zones: a) a crust with a thickness
of a few tens of kilometers, b) an upper mantle extending down from the
crust to about 1000 km, c) a lower mantle down to about 2900 km, d)
an outer core between 1250 km and 3480 km from the center, and e) an inner
core of radius $\approx 1220$ km. The density changes abruptly 
between the inner and
outer core, and also at the border between the lower mantle and the
outer core. The positions of these abrupt changes 
 are known with an
accuracy of better than a percent from seismological data.

Table~\ref{table} compares the mass and moment of inertia that we have
computed for each of the six earth models with the measured
values. The recent models given in \cite{PREM} and
\cite{stacey} reproduce the total mass and moment of inertia with 
 excellent precision. The older models \cite{HB1,B497,B1,A2} give 
 slightly worse fits to the mass and moment of inertia.
The models listed in the third to sixth row in Table~\ref{table} are
 in conflict with seismological measurements. 

The set of
 six models represents a sample that allows for variations of the
density distribution larger than the uncertainties in the PREM model.
As shown in Sec.~\ref{moments} (and the last four columns of Table~\ref{table}), the large differences
between the six density models produce relatively small 
(but not always negligible, see Sec.~\ref{moments})
changes in the
predicted characteristics of the earth regeneration effect for the
Super-Kamiokande, SNO, BOREXINO, ICARUS, and HERON/HELLAZ experiments.

 Figure \ref{density} shows that the largest differences
between the six models of the earth are in the core, below 2000
km. The operating solar neutrino experiments, and those currently
under development, are located at relatively high
northern latitudes.
Solar neutrinos that are detected in these experiments never cross the
inner core of the earth.  
 Among the real-time experiments that are currently
operating or are under construction, 
Super-Kamiokande is most sensitive to the core distribution. Nevertheless,
the fraction of
 a year during which the neutrinos cross the outer core at
the Kamioka site is small ($\simeq$ 7\%).

\subsection{Chemical Compositions}
\label{compositions}

Measurements of the propagation of seismological waves in the earth's
interior and  studies of the properties of minerals under high
pressure, have been combined to determine 
the chemical composition of the earth's interior with relatively high
accuracy \cite{corecomp,mantlecomp}.  Using the results of
references \cite{corecomp,mantlecomp},
we adopt
a best-estimate charge to mass ratio, Z/A, of 0.468 for the core (83\% Fe,
9\% Ni, and 8\% light elements with Z/A = 0.5) and 0.497 for the mantle
(41.2\% ${\rm SiO}_2$, 52.7\% ${\rm MgO}$ and 6.1\% ${\rm FeO}$).

A lower limit for the charge to mass ratio in the core is $0.465$, which
corresponds to assuming a composition of $100$\% iron. From the
seismic and mineral data,
geophysicists have concluded\cite{corecomp}
 that the minimum amount of iron in the
core is $80$\%.  We determine a maximum value of $Z/A = 0.472$ in the
core by assuming a composition of $80$\% iron and $20$\% light elements.
The total range of the electron number density 
 due to the imperfectly known 
composition of the core is about $1.5$\%. 

The chemical composition in the mantle is believed known to about
 $1$\% (see ref.~\cite{zharkov}).   We consider here variations of 
$-1$\% and $-2$\%.
 The value of 
$Z/X$ in the mantle cannot be increased 
significantly above the standard value of 
$0.496$ because that would require the presence of a large amount 
of hydrogen in the mantle.


\section{Average Event Rates and MSW Solutions}
\label{rates}

Including the earth regeneration effect, we have calculated
the expected one-year average event rates as functions of the neutrino
oscillation parameters $\Delta m^2$ and $\sin^22\theta$ for all four
operating experiments which have published results from their
measurements of solar neutrino event rates. These include the chlorine
experiment, Kamiokande, GALLEX and SAGE. Thus we update our previous
results given in Ref.~\cite{howell}, in which the earth effect was
neglected. We take into account, as before, the known threshold and 
cross-section
for each detector. In the case of Kamiokande, we also take into account
the known energy resolution (
20 \% $1\sigma$  at electron energy 10 MeV) and
trigger efficiency function \cite{Kam}.

We first calculate the one year average survival probability,
$\bar{P}_{SE}$, for a large number of values of $\Delta m^2$ and
$\sin^22\theta$ using the method described in Appendix A.  Then we
compute the corresponding one year average event rates in each
detector. We perform a $\chi^2$ analysis taking into account
theoretical uncertainties and experimental errors as described in
\cite{uncert}. 

Table \ref{snudata} summarizes the reported mean event
rates from each detector. We obtain allowed regions in $\Delta m^2$ -
$\sin^22\theta$ parameter space by finding the minimum $\chi^2$ and
plotting contours of constant $\chi^2 = \chi^2_{min} + \Delta\chi^2$
where $\Delta\chi^2 = 5.99$ for 95\% C.L. and 9.21 for 99\%
C.L.~\footnote{The C.L. in this paper are always for two degrees of
freedom. The values of $\chi^2$ are not reduced.}. 

The best fit is obtained for the small mixing angle (SMA) solution:

\begin{mathletters}
\label{SMAall}
\begin{eqnarray}
\label{SMApara}
\Delta m^2  &=&  5.0\times 10^{-6} {\rm eV}^2 ,\\ 
\sin^22\theta  &=&  8.7\times 10^{-3} ,
\label{SMAparb}
\end{eqnarray}
\end{mathletters}
which has a $\chi^2_{\rm min} = 0.25$.  
There are two more local minima of $\chi^2$.  The best fit for the
well known large mixing angle (LMA) solution occurs at
\begin{mathletters}
\label{LMAall}
\begin{eqnarray}
\label{LMApara}
\Delta m^2  &=&  1.3\times 10^{-5} {\rm eV}^2 ,\\
\sin^22\theta  &=& 0.63 ,
\label{LMAparb}
\end{eqnarray}
\end{mathletters}
with $\chi^2_{\rm min} = 1.1$.  There is also a less probable
solution, which we refer to as the LOW solution (low probability, low
mass), at~\cite{krastev93,BaltzW3} 
\begin{mathletters}
\label{LOWall}
\begin{eqnarray}
\label{LOWpara}
\Delta m^2 & = &1.1\times 10^{-7} {\rm eV}^2 ,\\
\sin^22\theta & = & 0.83 .
\label{LOWparb}
\end{eqnarray}
\end{mathletters}
with $\chi^2_{\rm min} = 6.9$. 
 The LOW solution is acceptable only
at 96.5\% C.L.

Figure~\ref{allowed} shows the allowed regions in  
the plane defined by $\Delta m^2$ and $\sin^2 2 \theta$.  The C.L. 
is 95\% for
the allowed regions of the SMA and LMA solutions and 99\% for the LOW
solution. 
The black dots within
each allowed region indicate the position of the local best-fit point
in parameter space.
The results shown in Fig.~\ref{allowed}
were calculated using the predictions of
the 1995 standard solar model of Bahcall and
Pinsonneault\cite{BP95}, 
which includes helium and heavy element diffusion;
the shape of the allowed contours depends only slightly upon the 
assumed solar model(see Fig.~1 of ref.~\cite{howell}).

The results given here differ somewhat from those given earlier in 
ref.~\cite{howell}, both because we are now including the regeneration
 effect and also because we are using more recent experimental data
 for the pioneering solar neutrino experiments.
Comparing the results given in 
Eqs.~(\ref{SMAall})--(\ref{LOWall}) and
Figs.~\ref{allowed}
with the corresponding allowed regions obtained 
for the same input neutrino experimental data but without including 
the earth effect shows that  terrestrial regeneration  
changes only slightly the best-fit solutions for the SMA
solution ($\lesssim 5\%$ in $\Delta m^2$ and $\sin^2 2 \theta$) 
and the LMA
solution ($\lesssim 10\%$ in $\Delta m^2$ and $\sin^2 2 \theta$).  The
values of $\chi^2_{\rm min}$ are also not significantly affected.  The LOW
solution is acceptable only if the regeneration effect is included;
otherwise, the LOW solution is ruled out at 99.9\% C.L.

Figure~\ref{survival} compares the  computed survival probabilities
for the day (no regeneration), the night (with regeneration), and the 
annual average.
These results are useful in understanding the day-night shifts in the 
energy spectrum that are computed and discussed in Sec.~\ref{spectrum}.
The  results in the figure refer to a detector at the location of
Super-Kamiokande, but the differences are very small between the survival
probabilities at the positions of Super-Kamiokande, SNO, and the Gran
Sasso Underground Laboratory. 

\section{The Zenith-Angle Exposure Function and the Zenith-Angle
Distribution Function}
\label{exposure}

In this section,
we define and calculate the zenith-angle exposure
function and show how the exposure function can be distorted 
by oscillations and by neutrino-electron interactions in the earth 
into the zenith-angle distribution function.

The fractional number of neutrino events observed 
as a function of solar zenith
angle, $\alpha$ (see Fig.~\ref{geometry}), is, for standard neutrino
physics, determined only by the latitude of the neutrino detector. The
number of events is largest for zenith angles at which the sun spends
the most time in its apparent motion around the earth. In what
follows, we shall refer
to the normalized number distribution of events, $Y(\alpha)$, as the
``zenith-angle exposure function'';  this function 
describes the relative amount
of time the detector is exposed to the sun at a fixed zenith angle.
We present in this section the
zenith-angle exposure function calculated assuming massless neutrinos
(no oscillations) and detectors placed at Kamioka, Japan (Kamiokande
and Super-Kamiokande), Sudbury, Canada (SNO), and the  Gran Sasso
Underground Laboratory (GALLEX, ICARUS, BOREXINO and HERON/HELLAZ).
The exposure function for the Homestake detector has been calculated
by Cherry and Lande`\cite{earthreg}.

The zenith-angle distribution of events can be distorted by the earth
regeneration effect implied by MSW solutions, since at 
 solar zenith angles larger than
$90^0$ neutrinos pass through the earth on their
way to the detector and therefore $\nu_e$'s can be regenerated by
interactions with electrons in the earth's interior. 
The distorted distribution function, $f(\alpha)$, will be referred to 
as the
``zenith-angle distribution.''  The main goal of this paper is to 
calculate and analyze the  shape of $f(\alpha)$ predicted by 
different MSW
solutions and to estimate the sensitivity of detectors to 
the difference between 
$f(\alpha)$ and the zenith-angle exposure function, $Y(\alpha)$.

\subsection{Zenith-Angle Exposure Function}
\label{subexposure}

In order to represent accurately the zenith-angle exposure function,
we consider 360 angles, $\alpha_i$, separated by $0.5^\circ$ intervals
between  0 and $180^\circ$. The fraction of
time in a year during which the zenith angle,
$\alpha$, (see Fig.~\ref{geometry}) of the sun is close to $\alpha_k$
is proportional to:

\begin{equation}
Y^{'}(\alpha_k) = \sum_{i=1}^{N}\theta(\alpha(t_i) - \alpha_k)
\theta(\alpha_{k+1} - \alpha(t_i)) \left[1~{\rm AU}/R(t_i)\right]^{-2}\,.
\label{Yprim}
\end{equation}
Here $\alpha(t_i)$ is the solar zenith angle at time $t_i$, $N =
T/\Delta t$, $T$ is the duration of one calendar year, and $\Delta t$
is the time-step.  The function $\theta$ is the well known
step-function $\theta (x) = 0$,  $x < 0$ and $\theta(x) = 1$, $x \geq 0$.
The function $R(t)$ is the
instantaneous earth-sun distance, and $1$ AU  is the one year 
 average earth-sun distance 
for which solar neutrino fluxes are routinely calculated in
standard solar models 
(see Appendix~\ref{A:astron} for 
an explicit representation of the time-dependent earth-sun distance).

The sum in Eq.~(\ref{Yprim}) was computed by simulating the motion of
the sun on the celestial sphere during one calendar year using the
formulae in Appendix~\ref{A:astron}. Computations with different
time steps between a few seconds to a few minutes are practically
equivalent for calculating event rates in the various solar neutrino
detectors. 

The normalized zenith-angle exposure function, $Y(\alpha_k)$, is obtained
directly from Eq.~(\ref{LMAall}),

\begin{equation}
Y(\alpha_k) = Y^{'}(\alpha_k)/\sum_{i=1}^{N}Y^{'}(\alpha_i) .
\label{Y}
\end{equation}
Thus $Y(\alpha_k)$ is the fraction of the time during one calendar
year that the sun's zenith angle is within the interval $(\alpha_k -
0.25^0,\alpha_k + 0.25^0$).
If the solar neutrino flux is constant in time (no neutrino
oscillations occur), then the function $Y(\alpha)$ will be the
normalized angular distribution of events in the detector. 
We use the analytical expressions for $R(t)$ given in
Ref.~\cite{almanac} (see Eq.~\ref{Rdep}).
The zenith angle exposure function was calculated analytically in 
ref.~\cite{lisimon}, without including the variation due to the changing
earth-sun distance.

Figures \ref{Adist} and \ref{AdistGS} show the undistorted zenith-angle 
exposure
functions for Super-Kamiokande, SNO, and, the detectors assumed 
to operate at 
Gran
Sasso (ICARUS, BOREXINO, and   
HERON/HELLAZ).  
Convenient numerical tables for the
$Y(\alpha_k)$ are available at http://www.sns.ias.edu/$^\sim$jnb.
Table \ref{latitudes} lists
the latitudes (all northern) of each of the solar neutrino detectors.

The positions of the
sharp peaks in  Figs.~\ref{Adist} and \ref{AdistGS}
are determined by the location of the
detector and by the obliquity, $\epsilon$, 
of the earth's orbit (approximately
$23.4^0$). The absolute value of the 
 difference between the maximum (or minimum) possible
zenith angle for a given location and the position of the closest peak
is equal to $\phi ~+~ \epsilon$.
At summer solstice,
the sun's zenith angle changes from the minimum possible to the angle
corresponding to the second peak (at an angle $> 90^0$). At winter
solstice the sun goes between the peak at angle $<90^0$ and the
maximum possible zenith angle. Thus during winter solar neutrinos pass
closer to the earth's center, whereas during summer they go through
lower density layers of the mantle.

\subsection{Zenith-Angle Distribution Function}
\label{distribution}

We calculate the event rates,
$Q_i$, along each direction, $\alpha_i$, with the aid  of a set of
survival probabilities, $P^i_{SE}$, computed (just once for each
direction, mixing angle, and $E/\Delta m^2$) along a fixed set of trajectories 
through
the earth.  The numerical value of each $Q_i$ is obtained from
Eq.~\ref{scrate} of Appendix C. The normalized zenith-angle distribution
function is

\begin{equation}
\label{andist}
f(\alpha_i) = Q(\alpha_i) Y(\alpha_i)/\sum_k Q(\alpha_k) Y(\alpha_k).
\end{equation}

In the absence of oscillations, the detector event rate, $Q_i$, is
independent of direction and disappears from the
right-hand side of the equation. In this case $f(\alpha) \equiv
Y(\alpha)$.

In Figs.~\ref{Adist} and \ref{AdistGS}, we present the expected distorted
angular distributions for the SMA, LMA, and LOW
solutions in Super-Kamiokande, SNO, ICARUS, BOREXINO and
HERON/HELLAZ. In the panels for Super-Kamiokande, SNO and ICARUS, we
show only the distributions corresponding to the best-fit points in
the SMA and LMA solutions; the curves corresponding to the LOW
solution are virtually undistorted.  Correspondingly, in the panels
for BOREXINO and HERON/HELLAZ, we show only the expected angular
distribution for the LOW solution (see Eq.~\ref{LOWparb}), 
since the SMA and LMA solutions
imply only a negligible distortion of the zenith-angle 
distribution for these low energy detectors.

\section{Moments of the Zenith-Angle Distribution}
\label{moments}

In this section, we evaluate   the expected  MSW
distortions of the angular distribution and compare the 
first two moments of the predicted  angular distribution of events 
 with the calculated 
 moments expected in the absence of oscillations. 
This comparison constitutes a new and, for the small mixing angle
solution, a more
powerful way of analyzing the time dependence of the observed neutrino
events. 
The predicted distortions of the recoil electron energy spectra in
Super-Kamiokande and in SNO were investigated in 
Ref.~\cite{moments,lisimon} 
in  terms of the analogous moments of the energy
distribution.

In Sec.~\ref{definitions}, we define  the first and second
moments of the zenith-angle distribution and then calculate in 
Sec.~\ref{predictionsofmoments} the predicted
MSW changes in the zenith-angle distributions.
We plot the  relative
shift of the first moment in the plane of the neutrino oscillation
parameters $\Delta m^2$ and $\sin^22\theta$, illustrating the
range of possible values of the shift of the first moment within the
95\% C.L. allowed  
 by the Chlorine \cite{homestake}, Kamiokande \cite{kamioka}, 
GALLEX\cite{GALLEX}, and SAGE\cite{SAGE} experiments. 

How sensitive will the 
Super-Kamiokande, SNO, ICARUS,
BOREXINO and HELLAZ/HERON experiments be to the regeneration effect?
We answer this question in Sec.~\ref{predictionsofmoments}
by plotting 
the number of standard deviations each MSW solution is
separated from the no-oscillation solution
in the plane defined by the values of the first two moments.

\subsection{Moment Definitions}
\label{definitions}

The first two moments of the zenith-angle distribution are
defined by

\begin{mathletters}
\begin{equation}
\langle\alpha\rangle = \int\alpha f(\alpha)d\alpha ,
\end{equation}
and
\begin{equation}
\sigma^2 = \int (\alpha - \langle\alpha\rangle )^2f(\alpha)d\alpha .
\end{equation}
\end{mathletters}
The fractional shifts in these two moments are defined as
\begin{mathletters}
\begin{equation}
\frac{\Delta \langle\alpha\rangle}{\alpha_0} = ( \langle\alpha\rangle - \alpha_0)/\alpha_0,
\end{equation}
and
\begin{equation}
\frac{\Delta\sigma^2}{\sigma_0^2} = (\sigma^2 -
\sigma^2_0)/\sigma_0^2 , 
\end{equation}
\end{mathletters}
where $\alpha_0$ and  $\sigma_0^2$ are 
the moments of the zenith-angle exposure function
distribution (no oscillations) and 
$\langle\alpha\rangle$ and  $\sigma^2$
 are the moments
of the distorted (by oscillations) zenith-angle distribution.
Because of the symmetry of $Y(\alpha)$ about $\alpha = 90 \deg$, 
$\alpha_0 = \pi/2$ for all the detector locations.
The values of the undistorted second moment, $\sigma_0$, are:
$37.9^0$ (Super-Kamiokande), $32.9^0$ (SNO), $34.9^0$ (Gran Sasso), 
$33.9^0$ (Homestake), $34.5^0$ (Baksan), and $47.5^0$ (Equator).

The Kamiokande collaboration excluded 
a significant region in $\Delta m^2$ - $\sin^22\theta$ parameter space
by grouping 
the events from below the horizon into five bins, 
corresponding to different zenith angles \cite{kamioka}. Lisi and
Montanino in Ref.~\cite{lisimon} calculated binned angular
distributions for Super-Kamiokande and SNO for different values of
$\Delta m^2$ and $\theta^2$.

The calculation of the 
first two moments of the zenith-angle distribution is subject to fewer
complications than the more traditional method used in
references~\cite{kamioka,lisimon} of 
binning the data and doing a $\chi^2$ analysis. For binned data, the
{\it a priori} unknown normalization and the angular dependence of the
detector sensitivity are, for example, 
two of the aspects of neutrino experiments that
introduce 
bin-to-bin correlations which 
are often  difficult to estimate from the observations or to evaluate
including all of the relevant detector
characteristics. On the other hand, the calculation of the first
and second moments directly from the data is straightforward.

\subsection{Predicted MSW Moments}
\label{predictionsofmoments}

Figures~\ref{MSWshift} and~\ref{sigmaMSWshift} show 
the expected shifts in the first two
moments as a function of the neutrino oscillation parameters, $\Delta
m^2$ and $\sin^22\theta$. 
We plot in Fig.~\ref{MSWshift} (Fig.~\ref{sigmaMSWshift}) 
contours of constant fractional shift,
$\Delta \alpha/\alpha_0$ ($\Delta \sigma^2/\sigma^2_0$),  
of the first moment (second moment) in the $\Delta m^2$ -
$\sin^22\theta$ plane. The five panels are, for both figures, 
for Super-Kamiokande, SNO,
ICARUS, BOREXINO and HERON/HELLAZ. For Super-Kamiokande, $\Delta
\alpha /\alpha_0$ varies 
between $-0.1$\% and 3\% in the SMA region and between
$0.5$\% and $7$\% in the LMA region. The corresponding ranges for
SNO are $(-0.1, 4.5)$\% (SMA) and $(0.3, 15)$\% (LMA). 
 SNO is somewhat more sensitive than
Super-Kamiokande to the shift of the first moment because $\nu_\mu$s
(and $\nu_\tau$s) do not contribute to the charged-current 
signal in SNO. ICARUS and
SNO have similar sensitivities to the regeneration effect since
both observe charged current reactions with high-energy thresholds
of 10.9~MeV and 6~MeV, respectively.  The low-energy experiments,
BOREXINO and
HERON/HELLAZ, are insensitive to both the SMA and LMA allowed regions,
but will be able to test the LOW solution to which neither SNO nor
Super-Kamiokande are sensitive.  The results for the second moment,
shown in Fig.~\ref{sigmaMSWshift},  exhibit similar trends.

Figure \ref{isosigma} 
summarizes the potential of the second generation of solar neutrino
experiments for discovering new physics via the earth regeneration
effect. The figure
displays iso-sigma ellipses, statistical errors only, in the plane of the
fractional percentage shifts of the first two moments, 
$\Delta \alpha/\alpha_0$ and $\Delta
\sigma^2/\sigma_0^2$. 
Assuming a total number of events of 30000, (which corresponds to 
$\sim 5$ years
of standard operation for Super-Kamiokande 
and $\sim 10$ years for SNO), we 
have computed the
sampling errors on the first two moments as well as the correlation of
the  errors using the following well known formulae
\cite{kendall}: $\sigma(m_1) = \sqrt{\frac{\mu_2}{N}}$, $\sigma(m_2) 
= \sqrt{\frac{\mu_4 - \mu_2^2}{N}}$, and $\rho(m_1, m_2) = 
\frac{\mu_3}{\sqrt{\mu_2}\sqrt{\mu_4 - \mu_2^2}}$.
Here $\mu_i$ is the $i-$th moment of the angular distribution of
events and $m_i$ are the corresponding estimates of the moments from a
sample of size N (number of events). The iso-sigma ellipses for the
 six detectors we consider here are centered around the undistorted
zenith-angle exposure function for which, by definition,
 $\Delta\alpha = \Delta\sigma^2 =
0$. Figure~\ref{isosigma} 
shows for each detector the predicted shifts of the
first two moments in the SMA, LMA, and LOW solutions.
The horizontal and vertical error-bars denote
the spread in predicted values of the shifts in the first two moments,
which are obtained by varying $\Delta m^2$ and $\sin^2 2\theta$ within
the 95\% C.L. allowed (see Fig.~\ref{allowed}) by the four pioneering
solar neutrino experiments.

For Super-Kamiokande (SNO), the current best-fit parameters 
$\Delta m^2$ and $\sin^22\theta$
  [see Eq.~(\ref{asym})] predict a
$5\sigma$ ($6.5\sigma$) effect for the SMA solution and $13\sigma$
($25\sigma$) effect for the LMA solution. 
Note that SNO is expected to require twice as much time to collect the
same number of events as Super-Kamiokande. In the same amount of
observing time, SNO and Super-Kamiokande are approximately equivalent
 for the SMA and Super-Kamiokande is significantly more efficient for the
LMA.

\section{Day-Night Asymmetry}
\label{ND}

  In this section, we calculate the day-night asymmetry 
\cite{earthreg,BaltzW,BaltzW2,BaltzW3,minn,lisimon,liu} caused by the 
earth regeneration effect. 
For Super-Kamiokande and SNO, the first accurate calculations 
of the  day-night asymmetry 
have been made only recently using 
a realistic density distribution
in the earth\cite{minn,lisimon}. 

We evaluate numerically the day-night asymmetry  for
Super-Kamiokande, SNO, ICARUS, BOREXINO, and HERON/HELLAZ. We present
quantitative estimates of the sensitivity of future experiments to the
predicted difference in night-time and day-time event rates.

The day-night asymmetry is defined as:

\begin{equation}
A_{n-d} = \frac{Q_n - Q_d}{Q_n + Q_d} ,
\label{asym}
\end{equation}
where $Q_n$ and $Q_d$ are respectively the average night-time and
day-time event rates during one  year. The calculation of
the asymmetry is made using the one-year average night-time and day-time 
 survival probabilities described in Appendix~A.

Figure \ref{MSWDNA} shows contours of constant day-night asymmetry in the
$\Delta m^2$ - $\sin^22\theta$ plane for Super-Kamiokande, SNO, ICARUS,
BOREXINO, and HERON/HELLAZ. The shaded regions in the panels for
Super-Kamiokande, SNO, and ICARUS 
are allowed (at 95\% C.L.) by the present data from Homestake,
Kamiokande, GALLEX and SAGE. In order to include the LOW solution near
$\Delta m^2 = 10^{-7}$ eV$^2$, the panels corresponding to BOREXINO
and HERON/HELLAZ show the allowed
regions at 99\% C.L.

Table \ref{magnND} gives the expected range of 
day-night asymmetries in Super-Kamiokande and SNO for both the SMA and
LMA solutions. The ICARUS sensitivity is similar to SNO.  
In general, SNO is slightly more sensitive than Super-Kamiokande 
for testing the day-night asymmetry because SNO detects a CC reaction;
 the sensitivity to NC scattering of muon neutrinos decreases somewhat
the sensitivity of the Super-Kamiokande detector. Both detectors are
 very sensitive to the large day-night asymmetry predicted in the
LMA solution. Their sensitivity to the SMA solution is smaller, 
especially in the low $\sin^22\theta$ corner ($\sin^22\theta < 5\times
10^{-3}$), where $A_{n-d} \simeq 0$. Both Super-Kamiokande and SNO are
 practically insensitive to the LOW solution because they can register
only high-energy boron neutrinos which do not resonate in the earth
for the low $\Delta m^2$ values in the LOW solution. As noted in
Ref.~\cite{BaltzW3}, $A_{n-d}$ is negative in a small region of parameter space 
for both Super-Kamiokande and SNO, i.e., the
night time event rate can be slightly lower than the day-time event
rate. 

The differences between SNO and ICARUS are mainly due to the
different assumed neutrino thresholds ($6.4$ MeV and $10.9$ MeV,
respectively) and to the locations of the detectors. For large mixing angles, for
which 
the regeneration effect takes place mainly in the mantle, 
 the sensitivities of the two detectors are particularly similar. At small
mixing angles ($\sin^22\theta < 0.3$) and $5\times 10^{-6} < \Delta
m^2/{\rm eV}^2 < 10^{-5}$, ICARUS is slightly more sensitive to the
earth effect because it is located at a lower latitude than
SNO.

The three future detectors that will measure low-energy neutrinos,
BOREXINO ($^7{\rm Be}$), HERON ($pp$),  and HELLAZ ($pp$), will be
sensitive to the large day-night asymmetries predicted in the LOW solution, but
insensitive to the asymmetries predicted by the SMA and LMA
solutions. The range of day-night asymmetries expected for the LOW solution are
($16 \pm 5\%$) for HERON/HELLAZ and ($16 \pm 8\%$) for BOREXINO.

\section{Which Statistical Test Is Best?}
\label{besttest}

Which statistical tests are most powerful in detecting new physics?
What type of analysis will most clearly show 
departures from the
zenith-angle exposure function due to the regeneration effect?  By
analyzing simulated data in this section, we shall see that the
preferred statistical analysis depends upon which 
solution Nature has chosen.

Table \ref{compare} compares the sensitivity of Super-Kamiokande
and SNO to the earth regeneration effect for three different
statistical tests.  We have computed the number of standard deviations
by which the best-fit MSW solutions (described in Sec. \ref{rates}) differ
from the undistorted zenith-angle exposure function. 
 We consider the first and second moments of the zenith-angle
 distribution (see Section \ref{exposure}), 
the  day-night asymmetry
($A_{n-d}$) (see Sec. \ref{moments}), and the 
Kolmogorov-Smirnov test of the
 distorted zenith-angle distribution. We assume 30000 events are detected
in the case of the SMA solution.  The comparison is made after only
5000 events are observed for the more-easily recognized LMA solution.

For the SMA solution, the moments
analysis is  most sensitive. The 
difference for SNO, between $4.9 \sigma$ (day-night analysis)
 and $6.5 \sigma$ (moments distribution), corresponds to 20000 events, 
 or approximately 7 years of data taking. All three statistical tests
can easily reveal the best-fit LMA solution, although the day-night
asymmetry is the most efficient characterization in this case.  The
Kolmogorov-Smirnov test is the least sensitive to the SMA solution, but
performs better than the moments method for the LMA solution.

We can understand physically why the SMA distortion 
is most easily detected by
measuring the moments while the LMA distortion is most prominent in
the day-night asymmetry.  Figure~\ref{ratiodistortion}
shows for Super-Kamiokande 
the fractional distortion, $[f(\alpha) - Y(\alpha)]/Y(\alpha)$,
of the zenith-angle distribution for the best-fit SMA and LMA
solutions.  One can crudely approximate the distortions by, for the
SMA solution,  a
delta-function near the maximum allowed zenith angle and, for the LMA
solution, a step function near $\pi/2$. With these simple
approximations, one can show analytically that the first moment and the
day-night asymmetry have similar statistical power for the SMA
solution, and the second moment is more discriminatory than either the
first moment or the day-night asymmetry. The reason that the second
moment is so useful for the SMA solution is that in this case the
distortion mostly arises when the neutrinos pass through the core at
large zenith angles.  Because the vacuum mixing angle is small, the 
enhanced mixing $[\rho_{\rm res} = 7 {\rm ~g~cm^{-3}}(E/10~{\rm MeV})]$
due to the earth matter effect is particularly
significant when the neutrinos traverse the core.
Since the vacuum mixing angle is large for the LMA, the matter
enhancement is not especially  significant in this case and the main
regeneration for the LMA solution is due to oscillations that occur in
the mantle, i.e., whenever $\alpha > \pi/2$.  The day-night asymmetry
is well-tuned to this distortion since $A_{\rm n-d}$ compares the
average event rate for $\alpha > \pi/2$ with the event rate for 
$\alpha < \pi/2$.  The LOW solution produces a relative distortion,
$[f(\alpha) - Y(\alpha)]/Y(\alpha)$, that has a shape similar to the 
LMA solution and is therefore most easily detected by the day-night
asymmetry. The Kolmogorov-Smirnov test is not optimally tuned to
any of the three best MSW  solutions and is therefore not as powerful
as the moments or the day-night asymmetry. 

\section{Moments of the Energy Spectrum}
\label{spectrum}

We refine in this section our previous calculations of the first two 
moments of the energy spectrum  from electron recoils produced by
interactions with $^8$B neutrinos. We  use  
here the slightly improved 
MSW solutions, described in Sec.~\ref{rates}, that include regeneration in
the earth. The reader is referred to our earlier paper\cite{moments} for 
the relevant definitions and notation (see also ref.~\cite{lisimon} for a
similar calculation). 

Table~\ref{fivemev} 
presents, for an assumed threshold of $5$ MeV, the first
and second moments of the electron recoil 
energy spectrum, and the percentage shifts 
with respect to the average electron  kinetic energy, $T_0$, and
the dispersion in the kinetic energy, $\sigma_0$, 
in the absence of oscillations.
The results for $T_0$ and $\sigma_0$ differ by small amounts ($<
1$\%) from our earlier results given in ref.~\cite{moments}; the present
results are numerically more precise.  
The results in Table~\ref{fivemev} 
are given for the best-estimate MSW solutions (SMA, LMA,
and LOW) described in Sec.~\ref{rates}.  For completeness, we list the 
one year average moments of the energy spectrum for day-time, 
night-time, and the total year.
The calculated moments for the Super-Kamiokande experiment are given
in the upper part of the table and the moments for SNO are listed in
the lower part of the table.
Table~\ref{sixmev} presents the same results for an assumed 
threshold of $6$ MeV.

We did not include in our calculation the unknown trigger efficiencies
of Super-Kamiokande or SNO.  The inclusion of these trigger functions 
can change the predicted first and second moments of the energy
distribution by a few percent and will certainly be included in the 
careful Monte Carlo calculations that will be performed ultimately by
the Super-Kamiokande and SNO experimental groups.

Comparing the calculated day and  night rates, Table~\ref{fivemev}
and Table~\ref{sixmev} show that 
regeneration in the earth 
slightly decreases , for both the SMA and the 
LOW solutions, the average kinetic energy of the recoil
electrons in both Super-Kamiokande and SNO. This decrease occurs
because in the sun these two solutions 
preferentially  transform low energy neutrinos
from $\nu_e$ to $\nu_\mu$ (or $\nu_\tau$) and therefore there is a
relatively larger chance at low energy of regenerating $\nu_e$ from
$\nu_\mu$ ( or $\nu_\tau$) in the earth.
For the LMA solution, regeneration increases the average kinetic
energy since in this case the high-energy part of the $^8$B neutrino
energy spectrum is preferentially depleted of $\nu_e$ in the sun.

The shift between day and night of the moments is most significant for
the LMA solution.
In fact, if the nature has chosen the LMA solution, then
the spectral distortion may be highlighted by
comparing the day-time and night-time moments.

\section{Sensitivity to Earth Models and Solar Models}
\label{modelsensitivity}

We calculate in Sec.~\ref{earthuncertainties} 
the sensitivity of the MSW predictions
to the assumed density profile and chemical composition of the earth model
and in Sec.~\ref{solaruncertainties} the dependence upon the assumed
model of the sun.

\subsection{Uncertainties Due to Earth Models}
\label{earthuncertainties}

Table \ref{table} presents the calculated percentage shifts of the
first two moments of the zenith-angle event distribution 
for all  six models of the earth discussed in 
Sec.~\ref{profiles}; the calculations were made assuming either the SMA or
the LMA solutions. The fractional changes of the first moment vary by
only $\sim 0.02$\% for the SMA solution and $\sim 0.2$\% for the LMA
solution, although the density profiles in some of these models are
significantly different from the range allowed by current
seismological data. We conclude that the shape of the zenith-angle 
 distribution can be calculated with acceptable accuracy for any of the
recently published density profiles of the earth.

Table \ref{chemical} illustrates the uncertainties
in the MSW predictions of the  shifts of
the first and second moments of the zenith-angle distribution
due to uncertainties
 in the electron number density in the mantle and in the
core. The ranges of Z/A included in the table ($\pm 2$\% in the core and
$-1$\%,
$-2$\% in the mantle) are larger than the current estimates of the 
geophysical
uncertainties (see the discussion in Sec.~\ref{profiles} and 
reference~\cite{corecomp,mantlecomp}).
 We conclude from Table~\ref{chemical} that uncertainties
in the chemical composition affect the predicted  moment shifts due to
regeneration by
at most a few percent of their values.

A simplified model with a uniform composition of $Z/A =
0.5$ has been used in reference~\cite{liu} 
(and  in many or most of the early
calculations related to the regeneration effect,  see~\cite{earthreg}).
The
predictions from this constant-composition model are also given in
Table~\ref{chemical}; this crude  model 
leads to imprecise, but not grossly erroneous, predictions of the 
moments of the zenith-angle
distribution.

\subsection{Uncertainties Due to Solar Models}
\label{solaruncertainties}

To the best of 
our knowledge, all previous discussions of the earth regeneration
effect have described this phenomenon as if it were completely
independent of solar models. This implicit assumption is not exactly
correct since the size of the earth regeneration effect depends upon
the flavor content of the incident neutrino beam, which must be
calculated by using a solar model to describe (for specified MSW
parameters) the production and conversion probabilities of $^8$B solar
neutrinos as a function of the  position in the sun at which the
neutrinos are created and the neutrino energy.
The slightly different density distributions in different solar models
have the largest effect, which is still quite small as we shall see
below,  on the inferred flavor content of the incident
solar neutrino flux.

In order to quantify the dependence of the predicted earth
regeneration effect upon the characteristics of the solar model, we
have calculated the fractional shifts of the first and second moments
of the zenith-angle event distribution using three different solar
models.  As our standard solar model, we adopt the model with helium and
heavy element diffusion of Bahcall and Pinsonneault~\cite{BP95}.
For comparison, we use the 1992 model of Bahcall and
Pinsonneault~\cite{BP92}, which includes helium diffusion (but not heavy
element diffusion) and somewhat less accurate input physics.  Finally,
we use the 1988 model of Bahcall and Ulrich~\cite{BU88}, which does not
include any diffusion and has less precise opacities, equation of
state,  and
other input data. 

Table~\ref{solar} shows that the MSW predictions are essentially
identical for 
the 1992 solar model with helium diffusion and the 1995 solar 
model with helium
and heavy element diffusion plus improved input data.
The 1988 solar model leads to predictions that can differ by as much
as 10\% for the SMA moments that will be measured by SNO.  However, 
this 1988 model is inconsistent with recent helioseismological
measurements since the 1988 model does not include
diffusion\cite{helio}.

We conclude that  predictions of the  earth regeneration effect are
practically independent of solar models as long as the models 
include diffusion (i.e., are consistent with helioseismology).

\section{Future Experiments at the Equator}
\label{equator}

Recently Gelb, Kwong and Rosen \cite{GKR} suggested building a new
detector similar to SNO close to the equator in
order to increase the sensitivity of the experiment to the earth
regeneration effect. An equatorial location maximizes the time
neutrinos pass through the core of the earth during one calendar
year. 

In this section, we calculate the size of the 
regeneration effect for hypothetical equatorial detectors and compare
with the sensitivity of the detectors in their actual positions. 
We consider equatorial analogues of the 
Super-Kamiokande, SNO, BOREXINO, and HERON/HELLAZ detectors.

Figure \ref{Adisteq} shows the predicted 
 zenith-angle
 distribution for detectors at the equator. The curve in the
upper left panel is the zenith-angle exposure function. The other five
panels show the distortion due to regeneration for equatorial
analogues of Super-Kamiokande, SNO, ICARUS,
 BOREXINO, and HERON/HELLAZ. For each detector, only the
predicted angular distribution functions are shown for the best-fit
solutions to which the relevant detector is sensitive: SMA and LMA for
the high-energy boron neutrino detectors (Super-Kamiokande, SNO
and ICARUS) and LOW for the low-energy neutrino detectors
(BOREXINO and HERON/HELLAZ).


Figure \ref{isosigmaeq} 
shows the iso-sigma ellipses for the four
equatorial detectors. BOREXINO and HERON/HELLAX would be more
sensitive to the LOW solution if these detectors were 
built at the equator. However, the high-energy neutrino detectors 
(Super-Kamiokande, SNO, and ICARUS) would remain insensitive to
the LOW solution even if they were moved to the
equator. Low-energy neutrino detectors (BOREXINO and
HERON/HELLAZ) also remain insensitive to the SMA and LMA
solutions even at the equator.

Table \ref{magnAnd0} shows the gain in sensitivity that would occur if
detectors like Super-Kamiokande and SNO were 
built at the equator. The enhancement is represented in the table by the
shift in the first and second moment of the zenith-angle 
distribution and by the day-night asymmetry. The enhancements would be
important for the best-fit SMA solution, but less significant for the 
 LMA solution. However, regions of parameter space in the LMA
solution for which the predicted shifts in the first and second moments
are small could be probed more precisely with detectors at the
equator.

\section{Discussion and Conclusions}
\label{discussion}

The conversion in the earth of $\nu_{\mu}$ (or $\nu_{\tau}$) to the more
easily detected $\nu_e$ is a distinctive
prediction of the MSW effect that offers the possibility of
unambiguously establishing the existence of physics beyond the
standard electroweak model.  Because of the importance of this
subject, we have carried out precise numerical calculations of the 
size of the regeneration effect predicted by different MSW parameters
that are consistent with the experimental results from the chlorine,
Kamiokande, GALLEX, and SAGE experiments.  Our results show the
potential of the new experiments,  Super-Kamiokande, SNO, ICARUS,
BOREXINO, HERON, and HELLAZ,  for discovering the regeneration
effect.

Our results provide the most precise predictions available of the 
expected zenith-angle distribution of the solar neutrino events in the
absence of new physics and in the presence of MSW distortions. 
The results are obtained by numerical
calculations that are discussed in Sec.~\ref{exposure} and
illustrated in Fig.~\ref{Adist} (for Super-Kamiokande and SNO) and
Fig.~\ref{AdistGS} (for the Gran Sasso experiments ICARUS, BOREXINO, and
HERON/HELLAZ).

We present the predictions for the small mixing angle
(SMA),
large mixing angle(LMA), and low mass (low probability, LOW) 
MSW solutions of
the solar neutrino problems. The parameters of these MSW solutions, which
are consistent with the results of the chlorine, Kamiokande, GALLEX,
and SAGE experiments, are given in Sec.~\ref{rates}.
Our solutions include self-consistently the effects of earth
regeneration.

Figure~\ref{allowed} shows  the allowed regions of the three MSW
solutions in the $\Delta m^2$ - $\sin^22\theta$ plane.
Figure~\ref{survival} presents the survival probabilities as a
function of energy for $\nu_e$ created in the sun.   This figure 
compares survival probabilities computed for the day (without
regeneration) with survival probabilities for the night (with
regeneration) and with  the average annual survival probabilities.

We describe 
the predicted MSW distortions in terms of the first
two moments of the zenith angle distribution of neutrino events (see Sec.~\ref{moments}), as
well as in terms of the traditional day-night asymmetry (see
Sec.~\ref{ND}). 
We analyze simulated data in Sec.~\ref{besttest} and show that the 
moments of the zenith-angle distribution are more sensitive to the
harder-to-detect SMA solution.  The 
predicted large effect of the LMA solution is more easily discovered
with the conventional day-night asymmetry.

The ``bottom line'' is illustrated succinctly in Fig.~\ref{isosigma}.
This figure shows that the current best-estimate MSW solutions predict
statistically significant deviations from the undistorted zenith-angle
moments for the Super-Kamiokande, SNO, and ICARUS experiments (which
are sensitive to the SMA and 
LMA solutions) and the
BOREXINO and HERON/HELLAZ experiments 
(which are sensitive to the LOW solution).

We have considered a number of effects that have not been previously 
investigated in connection with the earth regeneration effect. We have
calculated the sensitivity of the MSW predictions to a wide range of 
density profiles of the earth and also to a set of extreme chemical 
compositions. These calculations are discussed in
Sec.~\ref{profiles} and  
Sec.~\ref{earthuncertainties}.  We also evaluate the slight
dependence of the predicted earth regeneration effect upon the assumed
solar model used to calculate the flavor content of the incident
neutrino beam (see Sec.~\ref{solaruncertainties}).  Our results show
that these usually-neglected 
effects associated with the earth and solar models 
are rather small.

For completeness, we have carried out calculations for hypothetical
new detectors that might be built near the equator.  These
calculations are described in Sec.~\ref{equator} and show
quantitatively the enhanced sensitivity to the earth
regeneration effect of equatorial detectors, as emphasized by Gelb
{\it et al.}~\cite{GKR}. 

Using the best-fit MSW solutions calculated here that include the
earth regeneration effect, we have evaluated the first and second
moments of the electron recoil energy spectrum 
for $^8$B neutrinos detected in Super-Kamiokande and SNO.  
These calculations,
summarized in Sec.~\ref{moments} and in 
Table~\ref{fivemev} and Table~\ref{sixmev},
refine our earlier results\cite{moments} (see also ref.~\cite{lisimon})
for the moments of the electron recoil spectrum.
Perhaps most importantly, they show that for the LMA solution 
the comparison of the recoil electron energy
spectrum between day and night may reveal a distortion that is not
apparent in the temporal average of the energy spectrum.

What would we learn from an observation which showed that the neutrino
counting rate depended upon solar direction?
The experimental demonstration of a dependence of solar neutrino event
rate upon the direction of the sun would not only constitute a direct
proof of new physics but would at the same time eliminate a number of
the popular alternatives to the MSW effect.   Many of the 
alternatives to the MSW effect, such as vacuum oscillations, magnetic
moment transitions, and violations of the equivalence principle
predict that the counting rate is independent of the zenith angle
position of the sun.

\section*{Acknowledgments}
This work has been supported by NSF grant \#PHY-9513835. We are 
indebted to M. Fukugita, 
E. Lisi, and A. Smirnov for valuable comments on the draft
manuscript.  We are grateful to E. Lisi, 
W. Press,
P. Rosen, and A. Smirnov 
for stimulating discussions and to D. L. Anderson,
P. Goldreich, F. Press, 
and A. Rubin 
for valuable communications regarding seismological models of the earth.

\appendix
\section{NEUTRINO SURVIVAL PROBABILITIES}
\label{A:survP}

In order to calculate the event rates as a function of time we first
compute, following the prescription in Ref.~\cite{MikhS}, the electron
neutrino survival probabilities, $P_{SE}$, after traversing the
earth. We begin by using the analytical approximation
developed in Ref.~ \cite{KP} of the survival probabilities, $P_{S}$,
for an electron neutrino passing through the sun;  these solar
survival probabilities are averaged over
the relevant neutrino production regions for $^8{\rm B}$, $^7{\rm
Be}$, $pp$, $pep$, and CNO neutrinos. Assuming that the neutrinos
arriving at the earth represent an incoherent superposition of
mass-eigenstates\cite{MikhSm-uspekhi}, we calculate the electron
neutrino survival probability after passing through the earth,
$P_{SE}$, from the expression:

\begin{equation}
P_{SE} = {P_{S} - \sin^2\theta + P_{2e}(1 - 2P_S)\over \cos2\theta}.
\label{pse}
\end{equation}
Here $\theta$ is the neutrino mixing angle in vacuum and $P_S$ is the
probability of an electron neutrino produced in the sun to arrive as
an electron neutrino at the earth. The ambiguity in the above equation
for maximal mixing ($\cos2\theta = 0$) is only apparent. It can be
shown that in this case both $P_{S}$ and $P_{SE}$ are equal to
1/2. The probability $P_{2e}$ to find an electron neutrino after
passing through the earth, if the initial state describing a neutrino
entering the earth is the pure mass-eigenstate $\nu_2$, is calculated
numerically by integrating the evolution equations for neutrino states
in matter in the form given in \cite{MikhSmi-NuovoCim}. A key point in
the calculation is that these survival probabilities as a function of
$E/\Delta m^2$ need be calculated just once for each $\sin^22\theta$
and for a fixed set of trajectories. We have chosen a set
of 180 trajectories equally spaced between 90 and 180 degrees of
zenith angle separated by 0.5 degrees. This set is the same for each
solar neutrino detector since the density profiles used in our
calculations and described in Sec. \ref{profiles} are spherically
symmetric. The survival probabilities along each trajectory,
$P_{SE}^{i}$, have been calculated for a grid of 201x201 values of
$E/\Delta m^2$ and $\sin^22\theta$:

\begin{mathletters}
\label{neupar}
\begin{equation}
 10^{3}~{\rm MeV}/{\rm eV}^2 \leq E/\Delta m^2 \leq 10^{13}~{\rm
 MeV}/{\rm eV}^2 , \label{neupara}
\end{equation}
and
\begin{equation}
10^{-4} \leq \sin^22\theta \leq 1 .\label{neuparb}
\end{equation}
\end{mathletters}
The numerical precision of the calculated survival probabilities along
each trajectory is better than $0.1$\%.

The one-year averaged survival probability is given by:

\begin{equation}
\bar{P}_{SE} = \sum_{i=1}^{N} P_{SE}^i Y(\alpha_i) ,
\label{avsurvP}
\end{equation}
where the sum is over zenith angles from 0 to 180 degrees.  ($N = 360$
in our case). The zenith-angle exposure function is defined in
Sec. \ref{exposure}. Correspondingly, the one-year averaged
night-time survival probability is given by

\begin{equation}
\bar{P}_{SE}^{n} = \sum_{i=N/2}^{N} P_{SE}^i Y(\alpha_i) ,
\label{ntsurvP}
\end{equation}
where the sum now runs over angles from 90 to 180 degrees. Since the
night-time and day-time intervals within one year are equal, the
day-time event rate is simply

\begin{equation}
\bar{P}_{SE}^{d} = 0.5 P_{SE}.
\label{dtsurvP}
\end{equation}
With the calculated survival probabilities, it is straightforward to
determine the corresponding total event rates, as well as the day-time
and night-time one-year averaged event rates in any solar neutrino
detector.

\section{TIME DEPENDENCE OF THE ZENITH ANGLE}
\label{A:astron}

The dependence of the solar zenith angle on the time of the year and
on the  geographic
location of the detectors is given by the following set of 
formulae\cite{almanac}:

\begin{mathletters}
\label{Zangle}
\tighten
\begin{equation}
\cos\alpha = \sin\delta\sin\phi + \cos\delta\cos\phi\cos H\, , \label{zangla}
\end{equation}
\begin{equation}
\sin\delta = \sin\epsilon\sin\lambda\, , \label{zanglb}
\end{equation}
\begin{equation}
L = 280^0.461 + 0^0.9856003 n\, , \label{zanglc}
\end{equation}
\begin{equation}
n = -1462.5 + D + H\, , \label{zangld}
\end{equation}
\begin{equation}
g = 357^0.528 + 0^0.9856003 n\, , \label{zangle}
\end{equation}
and
\begin{equation}
\lambda = L + 1^0.915\sin g + 0^0.020\sin 2g\, . \label{zanglf}
\end{equation}
\end{mathletters}
The precision in the apparent coordinates of the Sun is $0^0.01$ and
the precision of the equation of time is 6 seconds between the years
1950 and 2050.  Here H is the fraction of day from $0^h$ UT, D is day
of the year (counting from January 1), n is the number of days from
Julian year 2000.0, $\lambda$ is the ecliptic longitude, $L$ is the
mean longitude of the sun (corrected for aberration), $\epsilon =
23^0.439 - 0.000 0004n$ is the obliquity of the ecliptic, $\delta$ is
the sun's declination, and $g$ is the mean anomaly\footnote{For
definitions of these astronomical quantities see Ref.~\cite{almanac}}.

The distance of the sun from the earth in astronomical units (1 AU =
$1.495978706(2) ~10^{11}$ m) is given by the formula:
\begin{equation}
R = 1.0014 - 0.01671\cos g  - 0.00014\cos2g .
\label{Rdep}
\end{equation}
Equation (\ref{Rdep}) has been used in the calculation of the day-night
asymmetries and the shifts of the first two moments of the
zenith-angle  distribution.

\section{CHARACTERISTICS OF FUTURE DETECTORS}
\label{A:futexp}

We describe in this Appendix the characteristics we have assumed for 
 Super-Kamiokande \cite{superK}, SNO\cite{sno}, ICARUS
\cite{icarus}, BOREXINO \cite{borexino}, HERON \cite{heron} and
HELLAZ\cite{hellaz}.  The Super-Kamiokande, SNO and ICARUS detectors
are sensitive only to high-energy $^8{\rm B}$ neutrinos, while
BOREXINO is sensitive primarily to $^7{\rm Be}$ neutrinos ($E_\nu =
0.862$ MeV) and the HERON and HELLAZ detectors are being designed to
detect low energy ($E_\nu < 0.44$ MeV) $pp$ neutrinos.

For the Super-Kamiokande detector, we adopt a threshold of 5 MeV and a
trigger efficiency of 50\% at this energy \cite{superK}. The energy
resolution function is assumed to have a gaussian shape with FWHM of
1.6 MeV at electron kinetic energy 10 MeV.  We have performed
calculations, which show that the sensitivity of our results to the
assumed energy resolution and trigger efficiency of the detector.

For the SNO detector, we calculate only the rate of the CC reaction,
namely $\nu_e + d \rightarrow p + p + e^-$. We adopt \cite{beier} a
threshold of 5 MeV and an energy resolution function with a $1\sigma$
uncertainty  of 1.1
MeV at 10 MeV electron kinetic energy.  The CC cross-section for SNO
was taken from Ref.~\cite{snoJE}. The trigger efficiency function has
been approximated with a step function at the threshold of the
detector.

Reference \cite{moments} gives further details regarding our
characterization of SNO and Super-Kamiokande.

For ICARUS, we have considered only the superallowed transition and
have used the neutrino absorption cross sections given in 
ref.~\cite{snp}. 
We have assumed a neutrino threshold for detection that corresponds to
electrons being produced with at least $5$ MeV of energy, which
requires a minimum neutrino energy of $10.9$ MeV. 

The BOREXINO detector is being developed as a 
neutrino-electron scattering experiment that will measure the flux of
$^7{\rm Be}$ neutrinos, 
using the step in the energy distribution of
neutrino-electron scattering events at the maximum recoil-electron
kinetic energy of $T_e = 0.62$ MeV. The detector
characteristics that are knowable {\it a priori} (which does not
include the crucial background rate versus energy) 
are not as important as for SNO and Super-Kamiokande
and we need to compute only the $\nu_e$ survival
probability. We include radiative corrections to the
neutrino-electron cross-section, calculated in \cite{BKS}, which for
recoil electron energies below 0.62 MeV 
are less than 1\%. The ratio of the ($\nu_\mu e$) to ($\nu_e e$) total
cross-sections for neutrino energy 0.862 MeV is
$\sigma_{\nu_\mu,e}/\sigma_{\nu_e,e} = 0.221$.

HERON and HELLAZ are also neutrino-electron scattering
experiments with very different preliminary designs, but with the same
target material, helium. They will measure the flux and spectral shape of
$pp$ neutrinos. We have assumed a threshold of 0.1 MeV 
 and a perfect energy resolution (delta function). The
trigger efficiency is represented by a step-function at the threshold
of the detector. We again use neutrino-electron cross-sections
including radiative corrections \cite{BKS}.

The event rate, $Q$, averaged over certain time interval, $\tau$, in a
neutrino-electron-scattering experiment, such as Super-Kamiokande,
BOREXINO, HERON or HELLAZ is given by

\begin{equation}
\langle Q\rangle_{\tau} = \int_0^{E_{\rm max}} \Phi(E_\nu) \left[ Z_{\nu_e}(E_\nu)
\langle P(E_\nu)\rangle_\tau~ + ~Z_{\nu_\mu}(E_\nu)\left( 1 -
\langle P(E_\nu)\rangle_\tau\right)\right] dE_\nu
\label{scrate}
\end{equation}

Here $Z_{\nu_e(\nu_\mu)}$ are the response functions of the detector
to either $\nu_e$ or $\nu_\mu$, $\Phi$ is the solar neutrino flux to
which the detector is sensitive, $E_\nu$ is the neutrino energy,
$E_{max}$ is the end point of the neutrino energy spectrum and
$\langle P\rangle_\tau$ is the average survival probability for the chosen time
interval $\tau$. The interval can be, e.g., the total day-time or
night-time during one calendar year, or a whole calendar year
including both days and nights. The calculation of the average
survival probabilities is described in Appendix~\ref{A:survP}
(Eqs.~\ref{pse}-\ref{dtsurvP}). The response functions represent the
convolution of the absorption cross sections with the detector 
characteristics (see Ref.~\cite{moments} for details).


\begin{table}
\caption[] {Sensitivity to the model of the earth.
The table illustrates the weak dependence on the model of
the earth of the calculated changes in the first and second moments of
the angular distribution of events in the Super-Kamiokande and SNO
detectors. The density distributions in the six models of the earth
listed in the table span a range of possibilities that is 
much larger than suggested by current
geophysical knowledge. The second and third columns give the total
mass, $M_\oplus$ (in $10^{27}$ g), and the moment of inertia, $I$ (in
$10^{45} {\rm g cm}^2$), for each model. 
The mass of the earth is $5.97370 \pm 0.00076$, the 
polar value of the moment of inertia is $0.804$,  and the
equatorial value is $0.801$~\cite{earththeory}.]
The last
four columns give the fractional  shift in percent 
of  the first two moments
($\Delta\alpha/\alpha_0$ and $\Delta\sigma^2/\sigma_0^2$) 
of the zenith-angle 
distribution of events in the Super-Kamiokande and SNO detectors for
the SMA (upper entry) and for the LMA (lower entry)
solutions.\protect\label{table}}

\begin{tabular}{l c c c c c c}
&&& $\Delta\langle\alpha\rangle/\alpha_0$&
$\Delta\sigma^2/\sigma_0^2$ 
& $\Delta\langle\alpha\rangle/\alpha_0$ & $\Delta\sigma^2/\sigma_0^2$ \\
Model & $M_\oplus$ & $I$ &Super-Kamiokande (\%)&Super-Kamiokande (\%)&SNO (\%)&SNO (\%)\\ \hline
PREM   & 5.972  & 0.802 & 1.02 & 2.03   & 1.04 & 2.42  \\
1981   &        &       & 3.23 & $-$0.33  & 5.28 & $-$2.20 \\ \hline
${\rm HB}'_1$ & 5.970 & 0.801 & 1.00   & 2.00 & 1.02 & 2.38 \\
1975   &        &       & 3.29 & $-$0.27  & 5.37 & $-$2.09 \\ \hline
B497   & 5.067  & 0.800 & 1.01 & 2.03   & 1.03 & 2.41 \\
1973   &        &       & 3.22 & $-$0.26  & 5.26 & $-$2.05  \\ \hline
B1     & 5.949  & 0.798 & 1.01 & 2.03   & 1.03 & 2.42 \\
1974   &        &       & 3.23 & $-$0.33  & 5.21 & $-$2.38  \\ \hline
${\rm A}''$     & 5.946 & 0.798 & 1.00   & 2.01 & 1.01 & 2.39  \\
1967   &        &       & 3.29 & $-$0.51  & 5.35 & $-$2.55 \\ \hline
Stacey & 5.973  & 0.802 & 1.00 & 2.01   & 1.02 & 2.38 \\
1969   &        &       & 3.27 & $-$0.28  & 5.34 & $-$2.11\\ 
\end{tabular}
\end{table}

\begin{table}
\caption[]{Solar neutrino data used in the analysis. The experimental
results are given in SNU for all of the experiments except Kamiokande,
for which the result is expressed as the measured $^8{\rm B}$ flux
above 7.5 MeV in units of ${\rm cm^{-2}s^{-1}}$ at the earth. The
ratios of the measured values to the corresponding predictions in the
standard solar model of Ref.~\protect\cite{BP95} are also given. The
result cited for the Kamiokande experiment assumes that the shape of
the ${\rm ^8B}$ neutrino spectrum is not affected by physics beyond
the standard electroweak model.\protect\label{snudata} }
\begin{tabular}{l c c c c c}
Experiment & Result & Theory & Units & Result/Theory & Reference \\
\hline

HOMESTAKE & $2.56 \pm 0.16 \pm 0.14$ &
$9.5^{+1.2}_{-1.4}$ & SNU & $0.27 \pm 0.022$ &\hfil\cite{homestake}\hfil\\

KAMIOKANDE & $2.80 \pm 0.19 \pm 0.35$ & $6.62 ~^{+0.93}_{-1.12}$ & $10^6$
cm$^{-2}$ s$^{-1}$ & $0.42 \pm 0.060$ &\hfil\cite{kamioka}\hfil \\ 

GALLEX & $69.7~\pm 6.7 ~\pm ~^{+3.9}_{-4.5}$ & $136.8^{+8}_{-7}$ & SNU &
$0.51 \pm 0.058$ & \hfil\cite{GALLEX}\hfil\\

SAGE & $72 ~^{+12}_{-10} ~^{+5}_{-7}$ & $136.8^{+8}_{-7}$ & SNU &
$0.53 \pm 0.095$ & \hfil\cite{SAGE}\hfil \\

\end{tabular}
\end{table}

\begin{table}
\caption[]{Northern latitudes (in degrees) for several solar neutrino
detectors. These angles correspond to $\phi$ in Fig.~\ref{geometry}.
\protect\label{latitudes}}
\begin{tabular}{l c c c c}
Homestake & Kamioka & Gran Sasso & Baksan & Sudbury \\ \hline
 44.3     &  36.4   & 42.4       & 43.3   & 46.3
\end{tabular}
\end{table}
\vskip 2cm

\begin{table}
\caption[]{Sensitivity to Solar Models. 
Values of the percentage  shifts of the first and
second moments of the zenith-angle event  distribution  in
Super-Kamiokande and SNO computed with three 
different solar models (BU1988,
BP1992, BP1995).\protect\label{solar}}
\begin{tabular}{l c c c c c}
&&\multicolumn{2}{c}{SMA} & \multicolumn{2}{c}{LMA}\\
Detector & Solar Model   & $\Delta\langle\alpha\rangle/\alpha_0$ & $\Delta\sigma^2/\sigma_0^2$ & 
$\Delta\langle\alpha\rangle/\alpha_0$ & 
$\Delta\sigma^2/\sigma_0^2$ \\ \hline
& 1995 & 1.02   & 2.03  & 3.22  & $-$0.32  \\
Super-Kamiokande &1992 & 1.02   & 2.03  & 3.22  & $-$0.32  \\
& 1988 & 0.96   & 1.94  & 3.22  & $-$0.32 \\ \hline
& 1995 & 1.04   & 2.42  & 5.28  & $-$2.20  \\
SNO &1992 & 1.04   & 2.44  & 5.28  & $-$2.20  \\
& 1988 & 0.96   & 2.26  & 5.27  & $-$2.20
\end{tabular}
\end{table}

\begin{table}
\caption[]{Day-night asymmetry in Super-Kamiokande and SNO. The table
gives the magnitude of the expected day-night asymmetry ($A_{n-d}$,
see Eq.~(\ref{asym})) (in percent) in Super-Kamiokande and SNO for
values of the neutrino oscillation parameters $\Delta m^2$ and
$\sin^22\theta$ corresponding to the best-fit SMA and LMA solutions
(see Eqs.~\ref{SMApara},\ref{SMAparb} and
\ref{LMApara},\ref{LMAparb}). The indicated uncertainties describe the
expected limits at 95\% C.L.\protect\label{magnND}}
\begin{tabular}{l c c}
Solution & Super-Kamiokande & SNO \\ \hline
SMA      & $1.8~~^{+4.2}_{-2.0}$  & $2~~^{+6.0}_{-2.4}$ \\
LMA      & $7.8~~^{+1.4}_{-6.6}$  & $15~~^{+29}_{-13}$
\end{tabular}
\end{table}

\vskip 1cm
\begin{table}
\caption[]{What statistical test is best? 
The table compares the  statistical power of three methods for 
analyzing data on the regeneration effect: 
a) moments of the zenith-angle distribution, b)
day-night asymmetry ($A_{n-d}$), and c) Kolmogorov Smirnov (K-S) test
of the zenith-angle distribution. 
The number of sigmas listed in the table corresponds to the deviation
of the best-fit MSW solutions described in  Sec. \ref{rates} from the
undistorted zenith-angle distribution.  The numerical results
correspond to 30000 events for the SMA solution and 5000 events for
the LMA solution.\protect\label{compare}}
\begin{tabular}{l c c c c}
&&Moments & $A_{n-d}$    & K-S test\\
Detector & Solution & ($\sigma$)&($\sigma$)&($\sigma$)      \\ \hline
Super-Kamiokande   & SMA      & $5$   & $4.4$    & $3.7$ \\
         & LMA      & $5.5$ & $9.2$  & $6.3$ \\ \hline
SNO      & SMA      & $6.5$ & $4.9$  & $4.4$ \\
         & LMA      & $10$  & $18.8$ & $12.4$
\end{tabular}
\end{table}

\begin{table}[htb]
\caption[]{Moments of the Energy Spectrum.
The moments in the absence of oscillations are 
$T_0 = 7.293$ MeV and $\sigma^2_0 = 3.391$ MeV$^2$ for
Super-Kamiokande and 
$T_0 = 7.646$ MeV and $\sigma^2_0 = 3.032$ MeV$^2$ for SNO.
\protect\label{fivemev}\ }
\begin{tabular}{lldddd}
\multicolumn{2}{l}{MSW Solution}&$\langle T\rangle$&$(T-T_0)/T_0$&$\sigma^2$
&$ (\sigma^2 - \sigma^2_0)/\sigma_0^2$\\
&&(MeV)&(\%)&${\rm (MeV^2)}$&(\%)\\ \hline
\multicolumn{2}{l}{\underline{Super-Kamiokande}}\\
\noalign{\medskip}
   &Day&     7.408& 1.58& 3.591& 5.88\\
SMA&Night  & 7.403& 1.50& 3.574& 5.38\\
&Average& 7.405& 1.54& 3.582& 5.64\\
\noalign{\medskip}
   &Day    & 7.275&\llap{$-$}0.25&3.368&\llap{$-$}0.67\\
LMA&Night  & 7.310& 0.23& 3.439& 1.42\\
&Average& 7.294&0.008& 3.408& 0.48\\
\noalign{\medskip}
   &Day &    7.298& 0.06&3.411& 0.58\\
LOW&Night  & 7.290&\llap{$-$}0.04& 3.398& 0.20\\
&Average& 7.294& 0.008& 3.404& 0.38\\
\noalign{\medskip}\hline
\noalign{\medskip}
\underline{SNO}\\
\noalign{\medskip}
   &Day    & 7.869&2.91&3.161& 4.25\\
SMA&Night  & 7.846& 2.62&3.144& 3.68\\
&Average& 7.858&2.76&3.152& 3.96\\
\noalign{\medskip}
   &Day    &7.644&\llap{$-$}0.03&3.032&\llap{$-$}0.03\\
LMA&Night  &7.719&0.95&3.113& 2.66\\
&Average& 7.687&0.53&3.084&1.72\\
\noalign{\medskip}
   &Day    &7.682&0.47& 3.069& 1.22\\
LOW&Night  & 7.668& 0.28& 3.060& 0.90\\
&Average& 7.675& 0.37& 3.064&1.06
\end{tabular}
\end{table}

\begin{table}[htb]
\caption[]{Moments of the Energy Spectrum.
The moments in the absence of oscillations are 
$T_0 = 8.057$ MeV and $\sigma^2_0 = 2.825$ MeV$^2$ for
Super-Kamiokande and 
$T_0 = 8.178$ MeV and $\sigma^2_0 = 2.348$ MeV$^2$ for SNO.
\protect\label{sixmev}\ }
\begin{tabular}{lldddd}
\multicolumn{2}{l}{MSW Solution}&$\langle T\rangle$&$(T-T_0)/T_0$&$\sigma^2$
&$(\sigma^2 - \sigma^2_0)/\sigma_0^2$\\
&&(MeV)&(\%)&${\rm (MeV^2)}$&(\%)\\ \hline
\multicolumn{2}{l}{\underline{Super-Kamiokande}}\\
\noalign{\medskip}
   &Day&      8.149&1.14& 2.968&5.07\\
SMA&Night  & 8.143&1.06& 2.954& 4.56\\
&Average& 8.146&1.10& 2.962& 4.84\\
\noalign{\medskip}
   &Day    & 8.046&\llap{$-$}0.14& 2.804&\llap{$-$}0.76\\
LMA&Night  &8.074& 0.22& 2.861& 1.25\\
&Average& 8.061&0.05& 2.833& 0.28\\
\noalign{\medskip}
   &Day &  8.064& 0.08& 2.836& 0.40\\
LOW&Night  & 8.058 & 0.01& 2.827& 0.048\\
&Average& 8.061&0.046& 2.832& 0.24\\
\noalign{\medskip}\hline
\noalign{\medskip}
\underline{SNO}\\
\noalign{\medskip}
   &Day    &  8.332& 1.88& 2.467 & 5.05\\
SMA&Night  &  8.315 &1.67& 2.452& 4.41\\
&Average&  8.323& 1.77 &2.459& 4.72\\
\noalign{\medskip}
   &Day    & 8.177&\llap{$-$}0.043& 2.347&\llap{$-$}0.043\\
LMA&Night  & 8.239& 2.63& 2.410& 2.63\\
&Average& 8.212&1.84& 2.392& 1.84\\
\noalign{\medskip}
   &Day    & 8.206& 0.34& 2.379& 1.28\\
LOW&Night  &  8.196& 0.22 & 2.370& 0.94\\
&Average& 8.201& 0.28& 2.374& 1.11
\end{tabular}
\end{table}

\begin{table}
\caption[]{Dependence of moments on assumed chemical composition. 
The table gives the relative shifts (in percent) 
of the first and second
moments ($\Delta\mu_{i}/\mu_i$) of the zenith-angle distribution in
Super-Kamiokande and SNO as a function of the assumed 
 chemical composition. The
ratio $Z/A$ in the core has been varied by $\pm 0.5$\%, $\pm 1$\%, 
$\pm 2$\%
(see second column) from the central value $\left[(Z/A)_{\rm core} =
0.465\right]$ 
adopted in the
rest of the paper. The ratio in the mantle has been varied by $-1$\%,
$-2$\% from the standard value of $\left[(Z/A)_{\rm mantle} = 0.496\right]$.  
The last row for each detector corresponds to a simplified model with
$Z/A = 0.5$ both in the mantle and in the core. The four columns for
each detector correspond to the first and second moment in the SMA and
LMA best-fit solutions respectively.\protect\label{chemical}}
\begin{tabular}{l c c c c c}
& &\multicolumn{2}{c}{SMA} & 
\multicolumn{2}{c}{LMA} \\
&$\Delta$(Z/A)/(Z/A)& $\Delta\langle\alpha\rangle/\alpha_0$ & $\Delta\sigma^2/\sigma_0^2$ & 
$\Delta\langle\alpha\rangle/\alpha_0$ & 
$\Delta\sigma^2/\sigma_0^2$ \\
Detector& (\%) & (\%)&(\%)&(\%)&(\%)\\ \hline
        & $-$2   & 1.03   & 2.06 & 3.22 & $-$0.338 \\
        & $-$1   & 1.02   & 2.05 & 3.22 & $-$0.329 \\
Super-Kamiokande  & $-$0.5 & 1.02   & 2.04 & 3.22 & $-$0.325 \\
&0.0&1.02&2.03&3.23&$-$0.323\\
        & +0.5 & 1.02   & 2.02 & 3.23 & $-$0.316 \\
(core)  & +1   & 1.01   & 2.01 & 3.23 & $-$0.311 \\
        & +2   & 1.01   & 1.99 & 3.23 & $-$0.301 \\ \hline
        & $-$1   & 1.01   & 2.03 & 3.19 & $-$0.336 \\
mantle  & $-$2   & 1.00   & 2.04 & 3.16 & $-$0.358 \\ \hline
Z/A = 0.5 & 0  &  0.98 & 1.89 & 3.27 & $-$0.257 \\ \hline
        & $-$2 & 1.05 & 2.46  & 5.27 & $-$2.23 \\
        & $-$1 & 1.04 & 2.44  & 5.27 & $-$2.22 \\
        & $-$0.5 & 1.04 & 2.43  & 5.27 & $-$2.21 \\
&0.0&1.04&2.42&5.28&$-$2.20\\
SNO     & +0.5 & 1.04 & 2.41  & 5.28 & $-$2.20 \\
        & +1   & 1.03 & 2.40 & 5.28 & $-$2.19 \\
        & +2   & 1.03 & 2.38 & 5.29 & $-$2.17 \\ \hline
        & $-$1   & 1.01 & 2.38 & 5.23 & $-$2.15 \\
mantle  & $-$2   & 0.98 & 2.33 & 5.18 & $-$2.10 \\ \hline
Z/A = 0.5& 0  &  1.01 & 2.28 & 5.33 & $-$2.11 \\ 
\end{tabular}
\end{table}

\begin{table}
\caption[]{Equatorial enhancement.  The magnitude of the regeneration
effect is compared for detectors located at their actual positions and
at the equator. The size of the effect is represented by the number of
standard deviations the first and second moments of the angular
distribution differ from the undistorted exposure function (and, in
parentheses, the day-night asymmetry in percent).  The values of the
neutrino oscillation parameters $\Delta m^2$ and $\sin^22\theta$
correspond to the best-fit SMA and LMA solutions (see
Eqs.~\ref{SMApara},\ref{SMAparb} and
\ref{LMApara},\ref{LMAparb}).  For comparison, the best-fit SMA and
LMA solutions produce $5\sigma$ and $13\sigma$ effects at the actual
location of Super-Kamiokande and $6.5\sigma$ and $25\sigma$ at the
actual location of SNO.\protect\label{magnAnd0}}
\begin{tabular}{l c c c}
Location   & Solution & Super-Kamiokande   & SNO \\ \hline
Equator    & SMA      & 8$\sigma$ ($3.0~~^{+5.0}_{-3.0}$ \%) &
$9\sigma$ ($4.1~~^{+7.9}_{-4.1}$ \%)\\
           & LMA      & 14.5$\sigma$ ($7.6~~^{+14}_{-6.3}$ \%) &
29$\sigma$ ($15~~^{+24}_{-13}$ \%)
\end{tabular}
\end{table}

\begin{figure}

\vskip 1cm
\caption[]{Schematic view of detector's location and sun's direction. 
 The zenith is defined as the line from the center
of the Earth through the center of the detector. The zenith angle, $\alpha$, 
and the
latitude of the detector's location, $\phi$, are also shown in the 
figure.\protect\label{geometry}}

\vskip 1cm
\caption[]{Density profiles for six different models of the earth. The
models are: 1) PREM\cite{PREM}, 2) Stacey's model \cite{stacey}, 3)
model HA\cite{A2}, 4) model ${\rm HB1}^{'}$ \cite{HB1}, 5) model
B497\cite{B497}, and 6) model B1 \cite{B1}. The radius is given in
kilometers from the center of the earth. All models are spherically
symmetric.\protect\label{density}}

\vskip 1cm
\caption[]{Allowed MSW solutions with regeneration.
The allowed regions are shown for the
neutrino oscillation parameters $\Delta m^2$ and $\sin^22\theta$.
The C.L. for the outer regions is 99\% 
 and the  C. L for the inner regions 
is 99\% (only applies to the LMA and SMA
solutions).
The data used here are
from the
chlorine\cite{homestake},Kamiokande\cite{kamioka},GALLEX\cite{GALLEX},
and SAGE\cite{SAGE} 
experiments.  The solar model used is the best standard model of
Bahcall and Pinsonneault (1995) with helium and heavy element
diffusion\cite{BP95}. 
The points where $\chi^2$ has a  local minimum are indicated
by a circle.
\protect\label{allowed}}

\vskip 1cm
\caption[]{Survival probabilities for MSW solutions.
The figure presents the survival probabilities for a $\nu_e$ created
in the sun to remain a $\nu_e$ upon arrival at the Super-Kamiokande
detector.  The best-fit MSW solutions including regeneration in the
earth are described in
Sec.~\ref{rates}.
The full line refers to 
the average
survival probabilities computed  taking
into account regeneration in the earth and the dotted line refers to
calculations for the day-time 
that do not include regeneration. 
The dashed line includes regeneration at night.
There are only slight
differences between the computed regeneration probabilities for
the detectors located at the positions of Super-Kamiokande, 
SNO and the Gran Sasso
Underground Laboratory.
\protect\label{survival}}

\vskip 1cm
\caption[]{Super-Kamiokande and SNO zenith-angle distributions. 
The figure shows the expected zenith-angle distribution of
 neutrino events during one calendar year in
 the Super-Kamiokande and the SNO detectors.  The angle, $\alpha$, 
 represents the angular separation between the direction to the sun
and the direction of the local zenith 
(see Fig.~\ref{geometry}). The two left panels display the zenith-angle
exposure functions, the undistorted angular distributions in the absence of
oscillations. The exposure functions are determined by the
location of the two detectors at, respectively, Kamioka, Japan, and
Sudbury, Canada.  The distorted zenith-angle distributions due
to the regeneration effect in the Earth are shown in the two right
panels; the neutrino solutions are indicated by: SMA (solid
line) and LMA (dotted line).\protect\label{Adist}}

\vskip 1cm
\caption[]{Gran Sasso zenith-angle distributions. 
The figure shows the expected zenith-angle distribution of
events during one calendar year in detectors
located at the Gran Sasso Laboratory in Italy: ICARUS, BOREXINO, HERON
and HELLAZ. The upper left panel shows the zenith-angle exposure
function, which does not depend on detector characteristics.
The
three additional panels display the distorted zenith-angle distributions
due to the regeneration effect in the Earth; the solutions are indicated by: 
SMA (solid line), LMA
(dotted line) and LOW (dashed line).
\protect\label{AdistGS}}

\vskip 1cm
\caption[]{Contours of constant relative shift (in percent) of the
average zenith angle, $(\langle\alpha\rangle - \alpha_0)/\alpha_0$, due to $\nu_e$
regeneration in the earth as a function of the neutrino oscillation
parameters, $\sin^22\theta$ and $\Delta m^2$. Here $\alpha_0 = 90^0$
is the average angle of the undistorted angular distribution with no
oscillations.  The shaded regions in the panels for Super-Kamiokande and
SNO are allowed by the latest solar neutrino data at 95\% C.L. and
represent the SMA and LMA solutions. In the lower two panels (BOREXINO
and HERON/HELLAZ) the three shaded regions are allowed at 99\% C.L.,
the low-mass region representing the LOW solution (see text for
details).
The black circle within each allowed region represents the point
which corresponds (locally) to the 
best-fit to the data.\protect\label{MSWshift}}

\vskip 1cm
\caption[]{Contours of constant relative shift (in percent) of the
dispersion of the 
 zenith angle, $(\sigma^2 - \sigma^2_0)/\sigma^2_0$, 
due to $\nu_e$
regeneration in the earth as a function of the neutrino oscillation
parameters, $\sin^22\theta$ and $\Delta m^2$. The values of $\sigma^2_0$
are given in the text for each of the experiments.  
The definition of the shaded regions is the same as in 
Fig.~\ref{MSWshift}.
\protect\label{sigmaMSWshift}}

\vskip 1cm
\caption[]{How many sigmas?  The figure shows the sensitivity of 
Super-Kamiokande, SNO, ICARUS, BOREXINO and
HERON/HELLAZ to the regeneration effect.  Iso-sigma contours,
statistical errors only, 
delineate the
fractional percentage 
shifts of the first two moments of the angular distribution
of events for an assumed 30000 observed events.  
For all but the ICARUS experiment,
the best-fit MSW solutions are indicated by black circles
(SMA), squares (LMA), and triangles (LOW);
the best-fit solutions are presented in Sec.~\ref{rates}.   The error
bars on the predicted moments correspond to $\Delta m^2$ and
$\sin^22\theta$ within allowed solution space  at 95\% C.L. (for
Super-Kamiokande, SNO, and ICARUS) or 99\% C.L. 
(BOREXINO and HERON/HELLAZ).
For ICARUS, we have indicated the best-fit solutions by a transparent
circle, square, or triangle. The best-fit SMA and LOW solutions for
ICARUS and the LOW solution for SNO are all three close together at
about $3\sigma$ from the no oscillation solution. In order to avoid
too much crowding in the figure, we have not shown the theoretical 
uncertainties for ICARUS.  
\protect\label{isosigma}}

\vskip 1cm
\caption[]{Contours of constant day-night asymmetry, $A_{n-d}$ (see
Eq.~(\ref{asym})), in Super-Kamiokande, SNO, ICARUS, BOREXINO and
HERON/HELLAZ.  The shaded regions are the same as in Fig.~\ref{MSWshift}. 
\protect\label{MSWDNA}}

\vskip 1cm
\caption[]{Relative distortion for Super-Kamiokande. The figure shows
the fractional distortion,
$[f(\alpha) - Y(\alpha)]/Y(\alpha)$, 
 of the zenith-angle distribution for the
best-estimate SMA and LMA MSW solutions.
\protect\label{ratiodistortion}}

\vskip 1cm
\caption[]{The zenith-angle distribution for equatorial
detectors with the characteristics of Super-Kamiokande, SNO, ICARUS,
BOREXINO and HERON/HELLAZ. Notation is the same as in
Figs.~\ref{Adist} 
and \ref{AdistGS}.\protect\label{Adisteq}}



\vskip 1cm
\caption[]{How many sigmas at the equator?  The figure shows the 
sensitivity to the regeneration effect of equatorial
detectors with the characteristics of Super-Kamiokande, SNO, ICARUS,
BOREXINO and HERON/HELLAZ. Notation is the same as in
Fig.~\ref{isosigma}.
\protect\label{isosigmaeq}}

\end{figure}

\newpage
\input psfig

\centerline{\psfig{figure=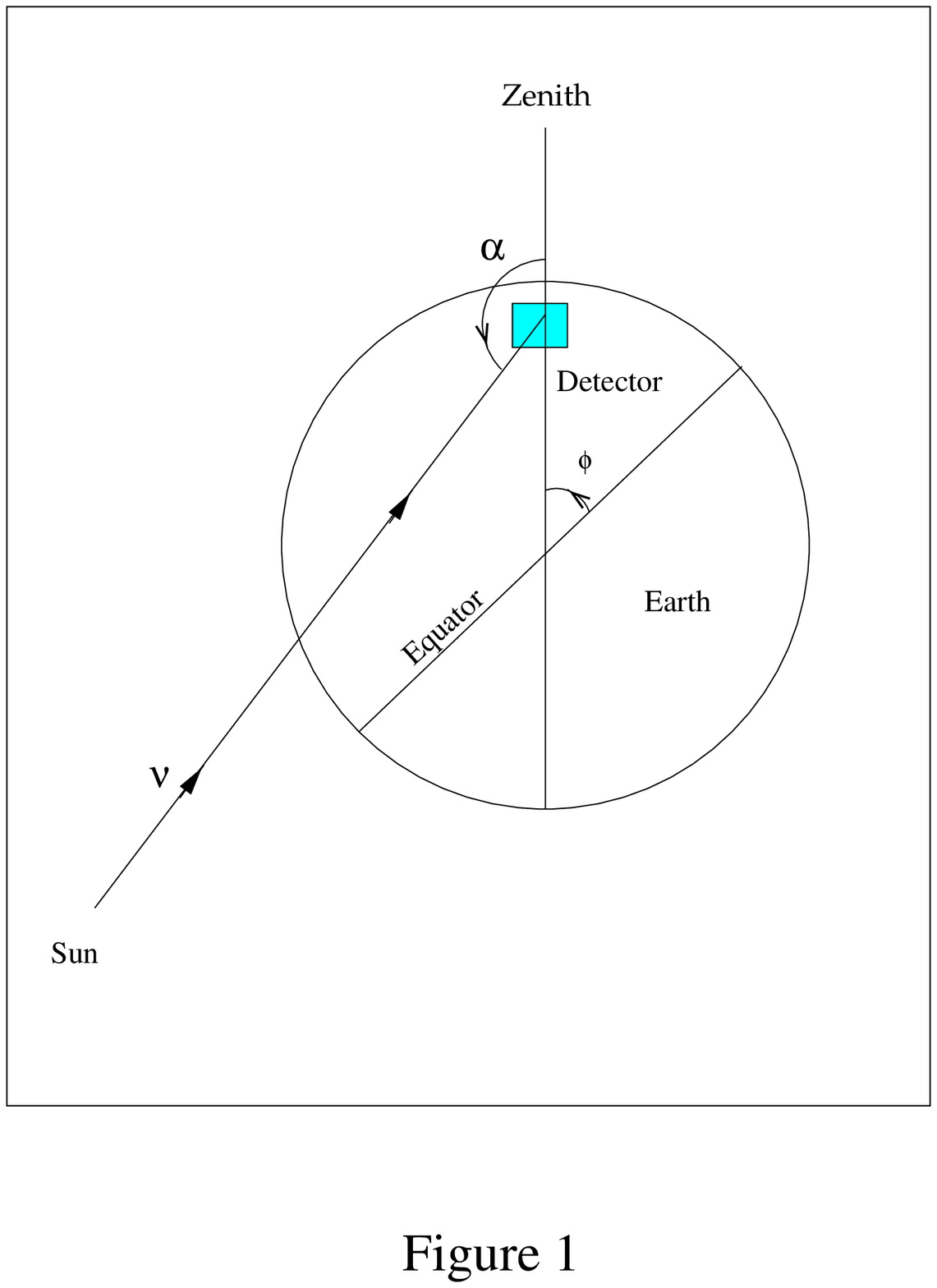,width=5.5in}}
\newpage
\centerline{\psfig{figure=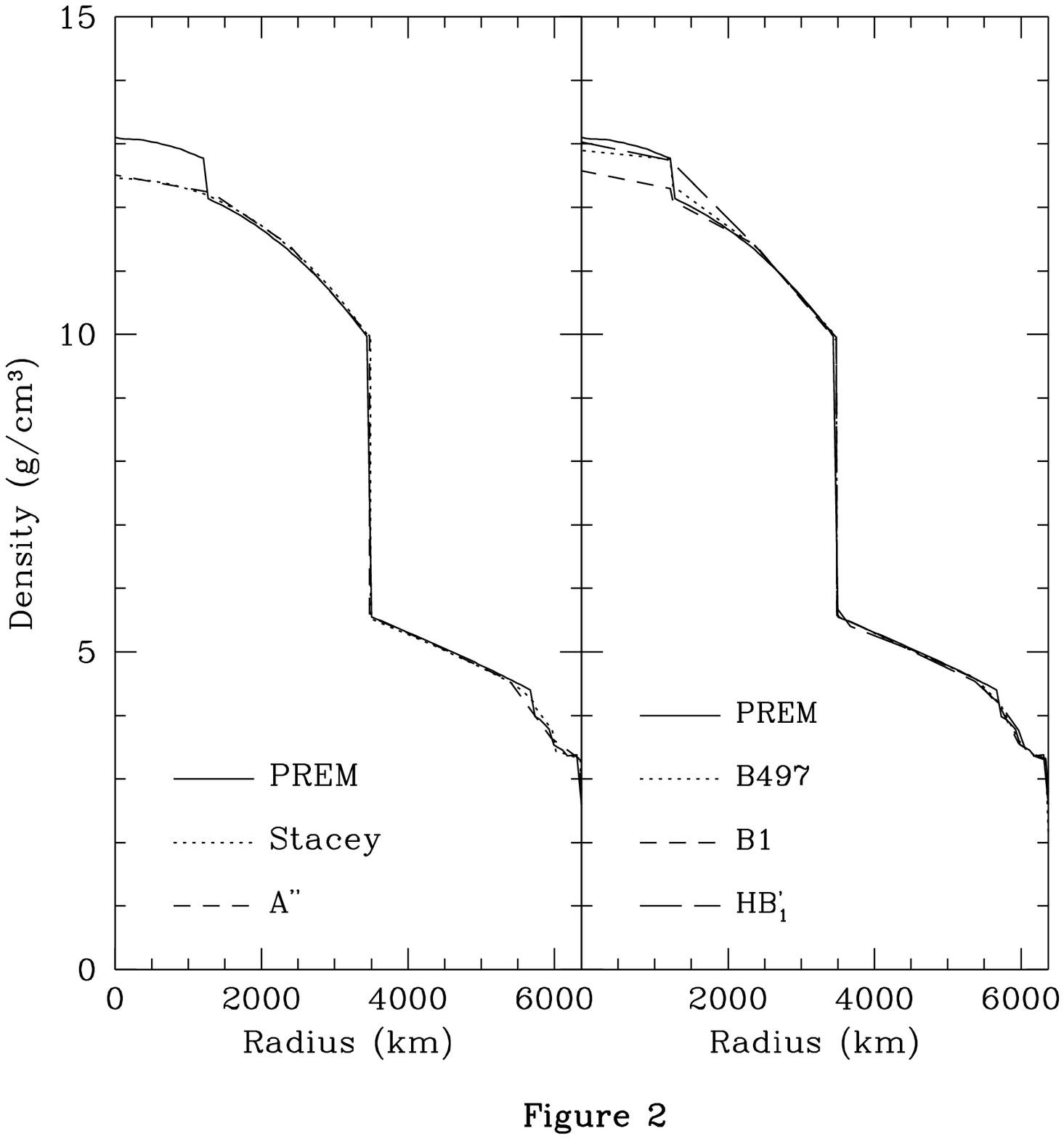,width=5.5in}}
\newpage
\centerline{\psfig{figure=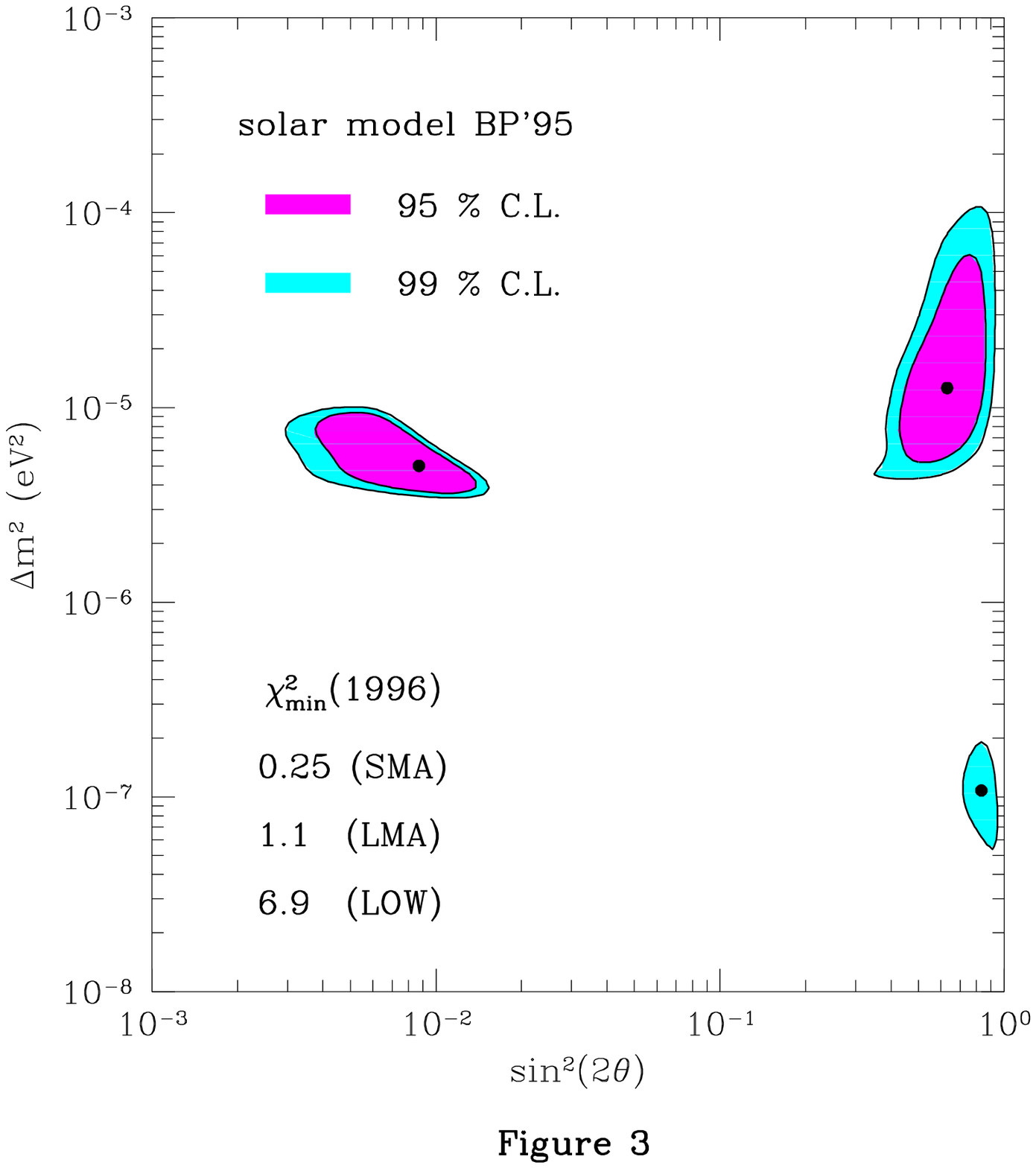,width=5.5in}}
\newpage
\centerline{\psfig{figure=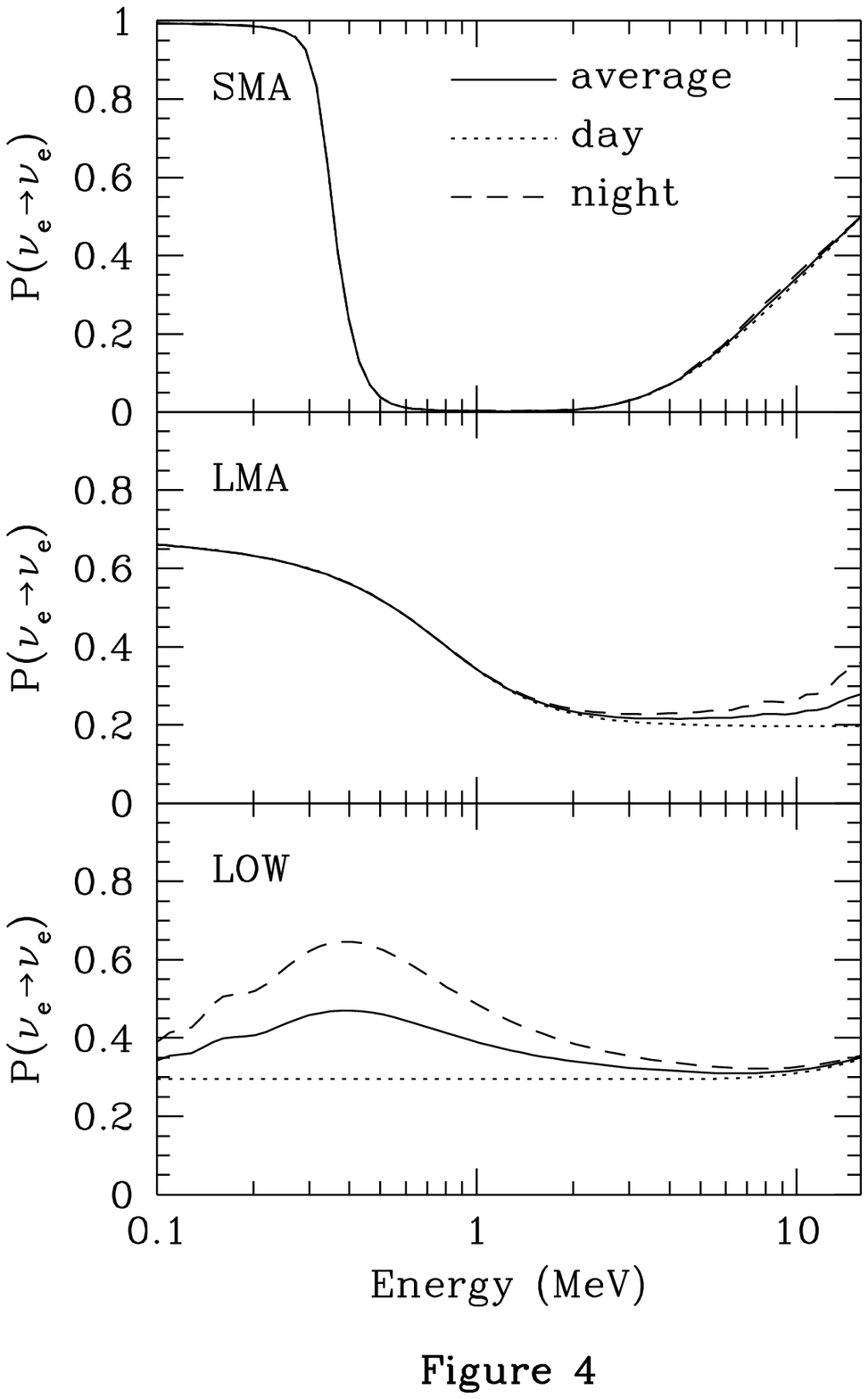,width=5.5in}}
\newpage
\centerline{\psfig{figure=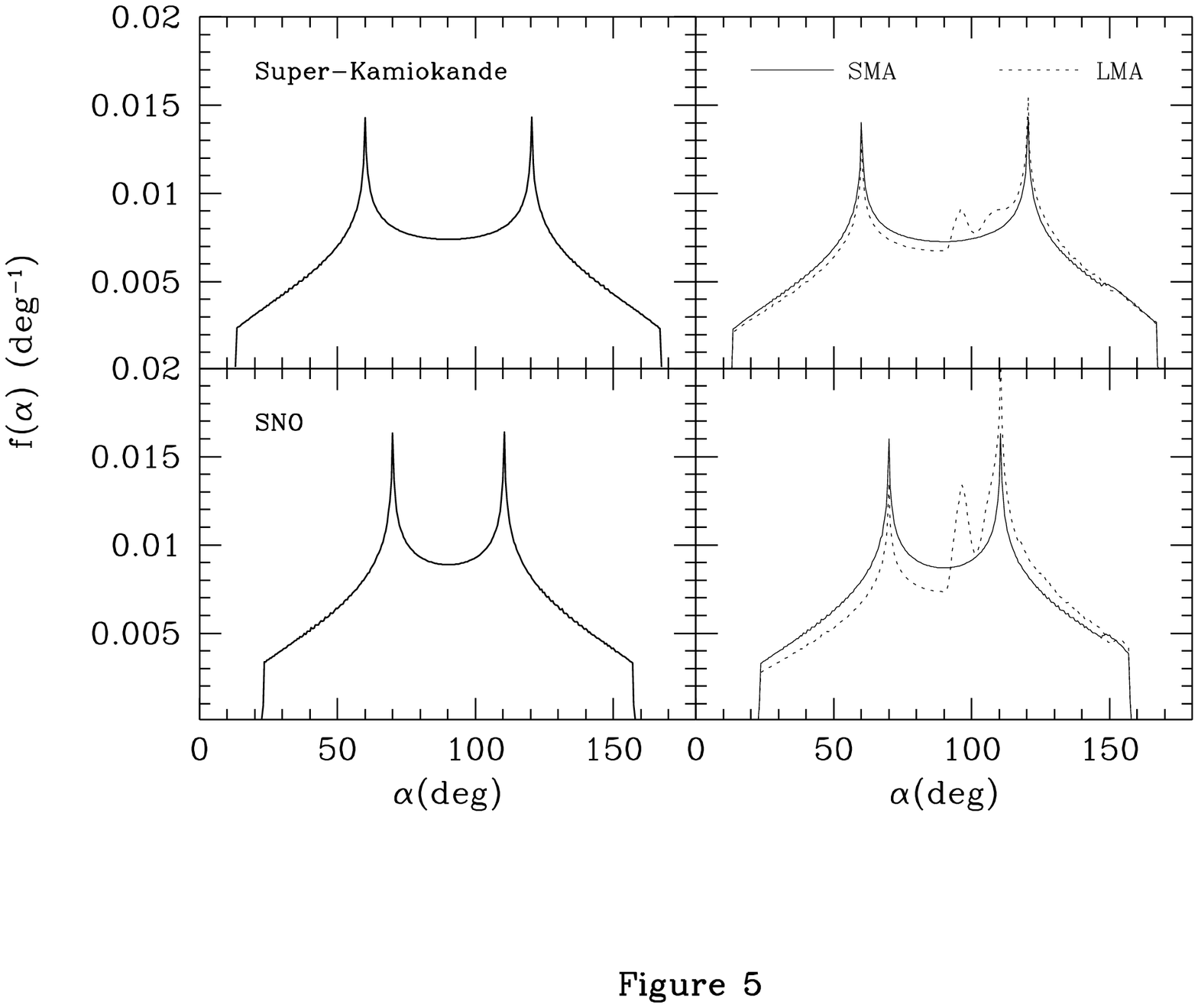,width=5.5in}}
\newpage
\centerline{\psfig{figure=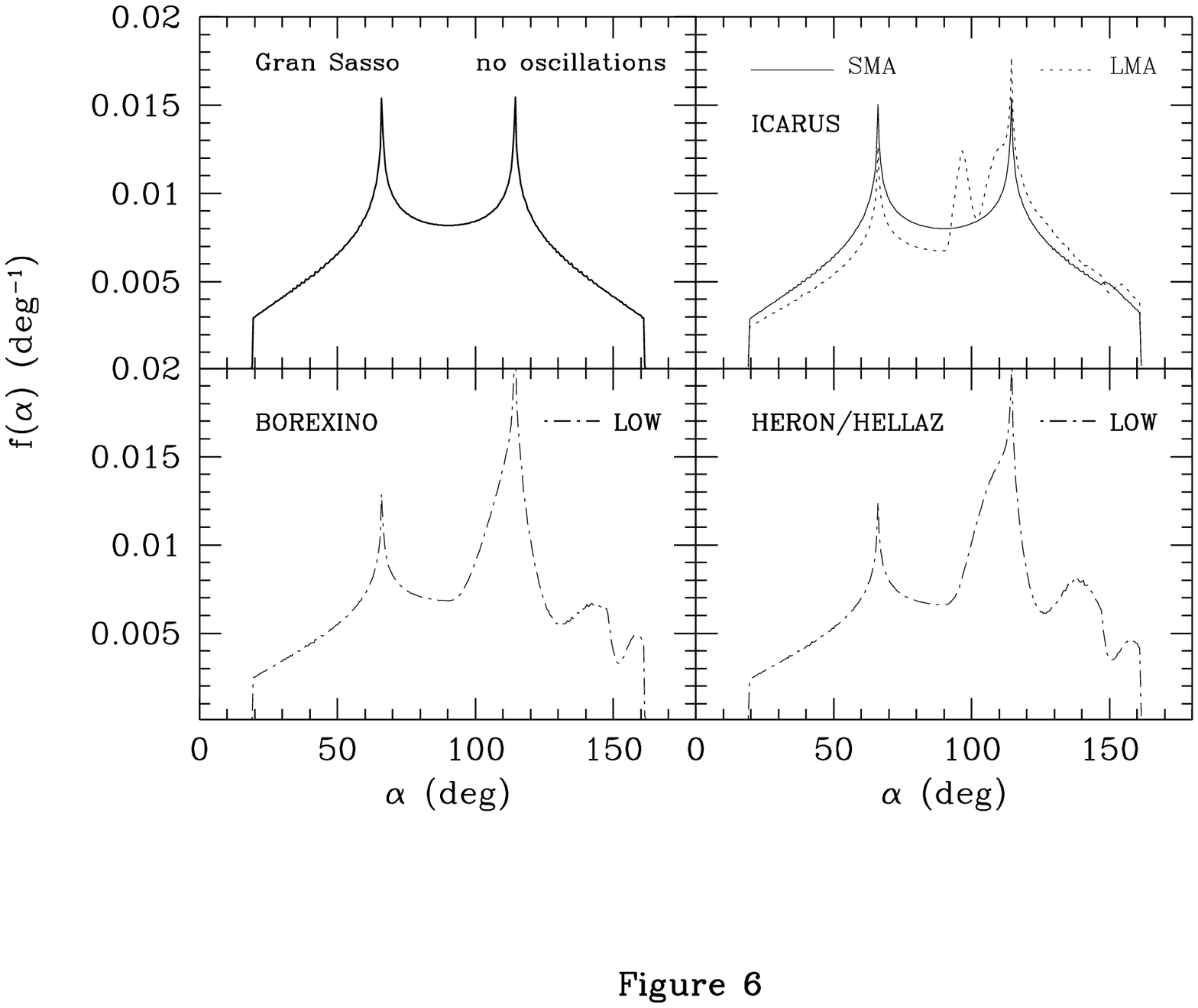,width=5.5in}}
\newpage
\centerline{\psfig{figure=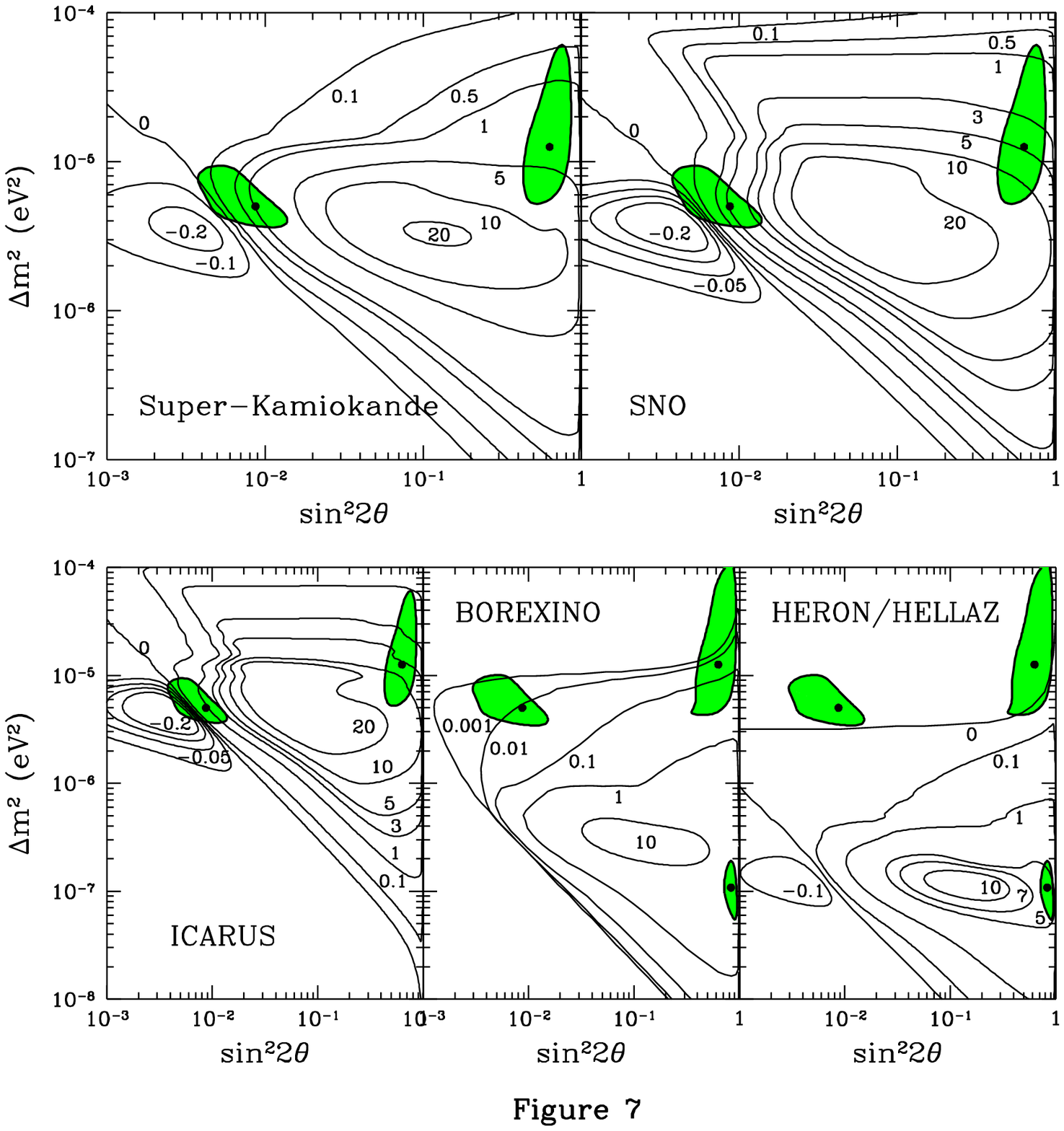,width=5.5in}}
\newpage
\centerline{\psfig{figure=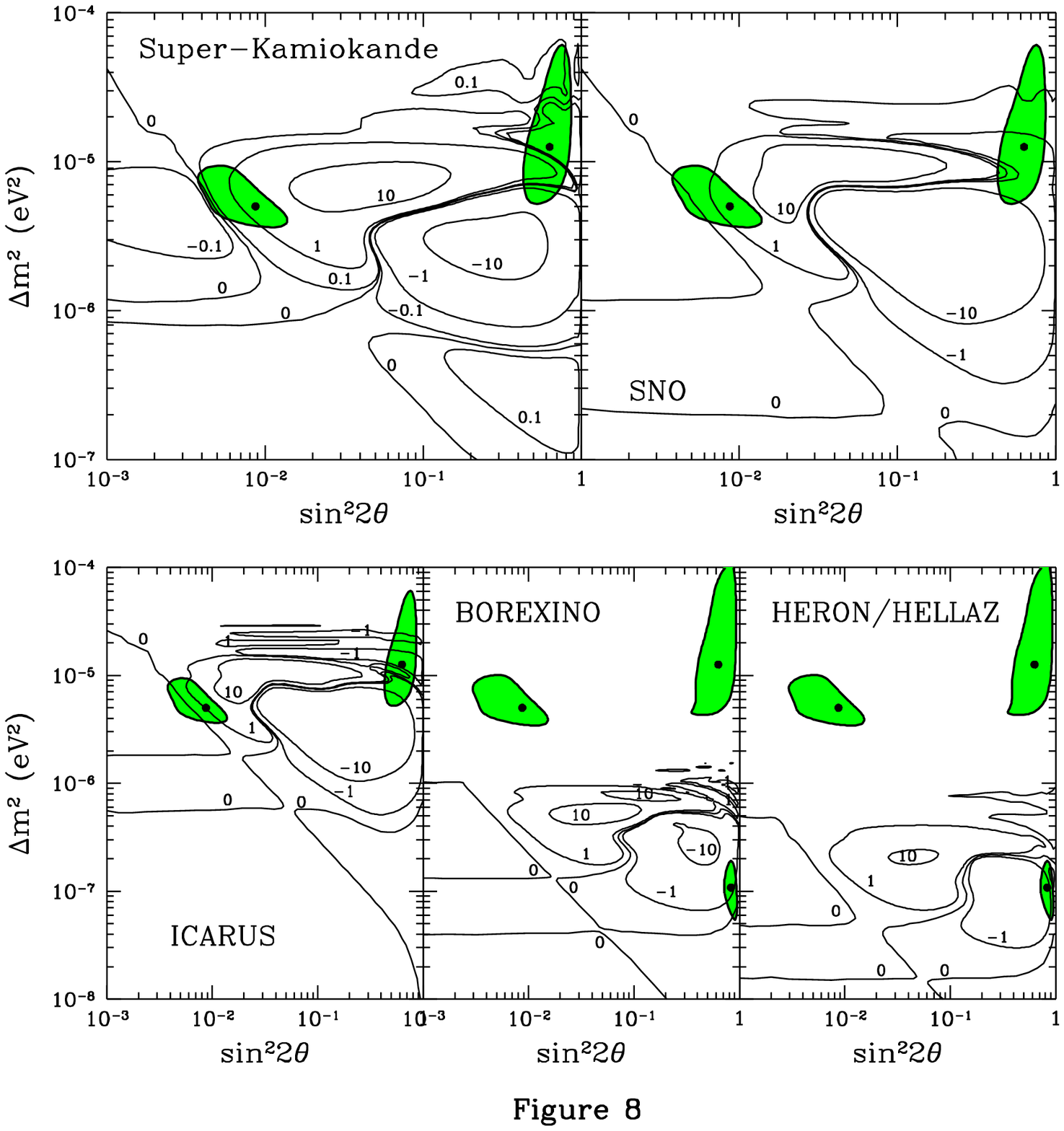,width=5.5in}}
\newpage
\centerline{\psfig{figure=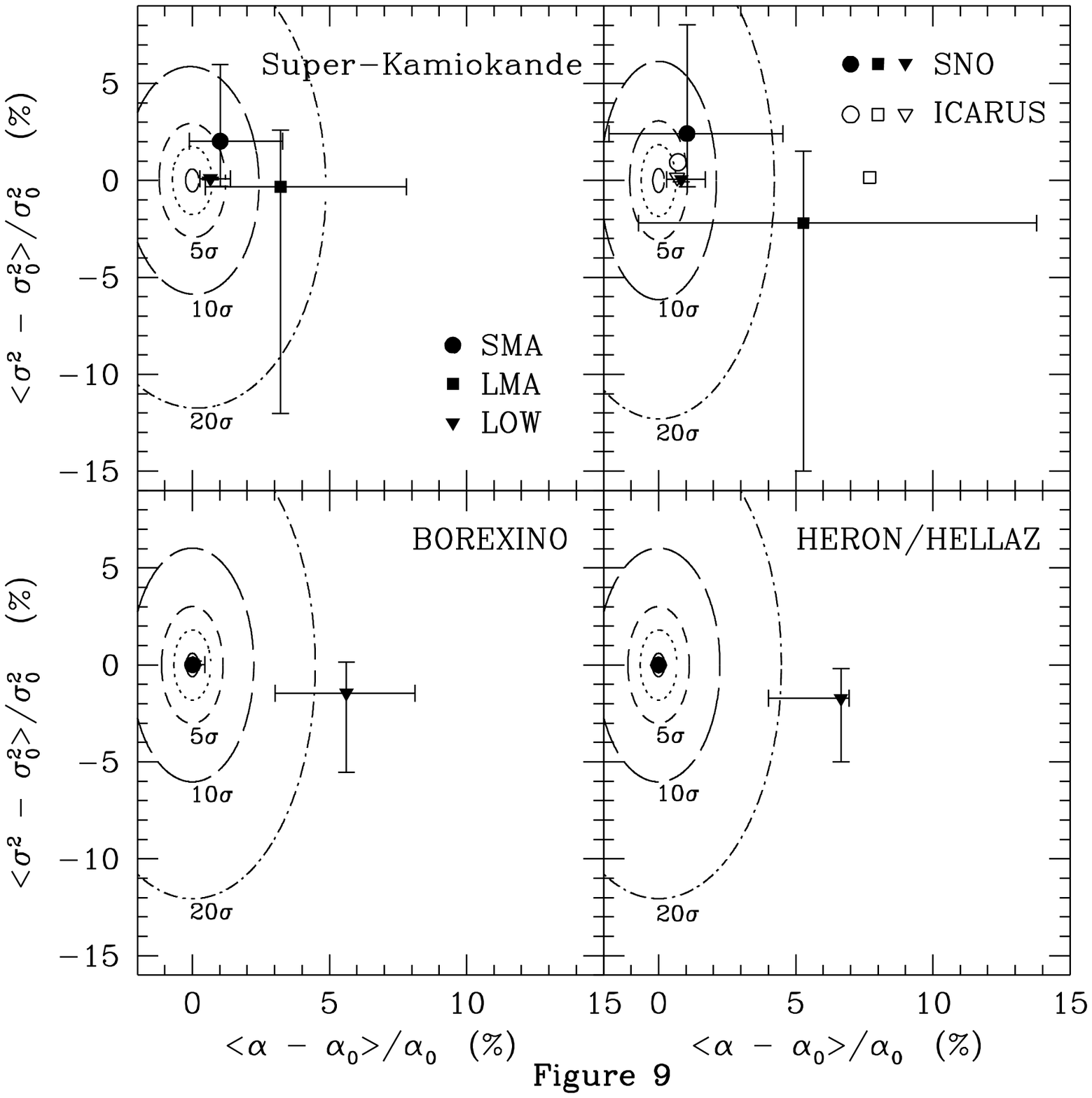,width=5.5in}}
\newpage
\centerline{\psfig{figure=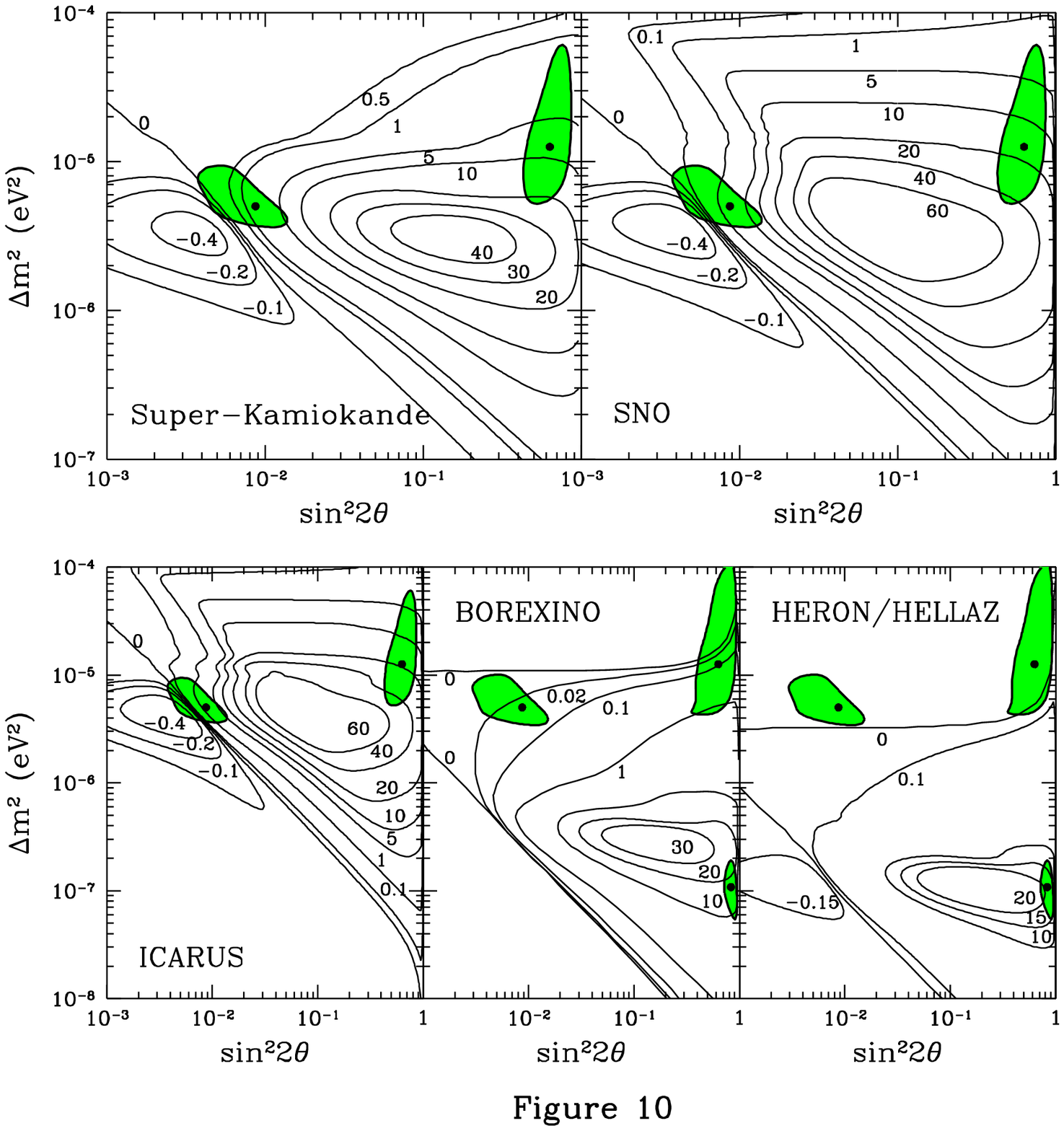,width=5.5in}}
\newpage
\centerline{\psfig{figure=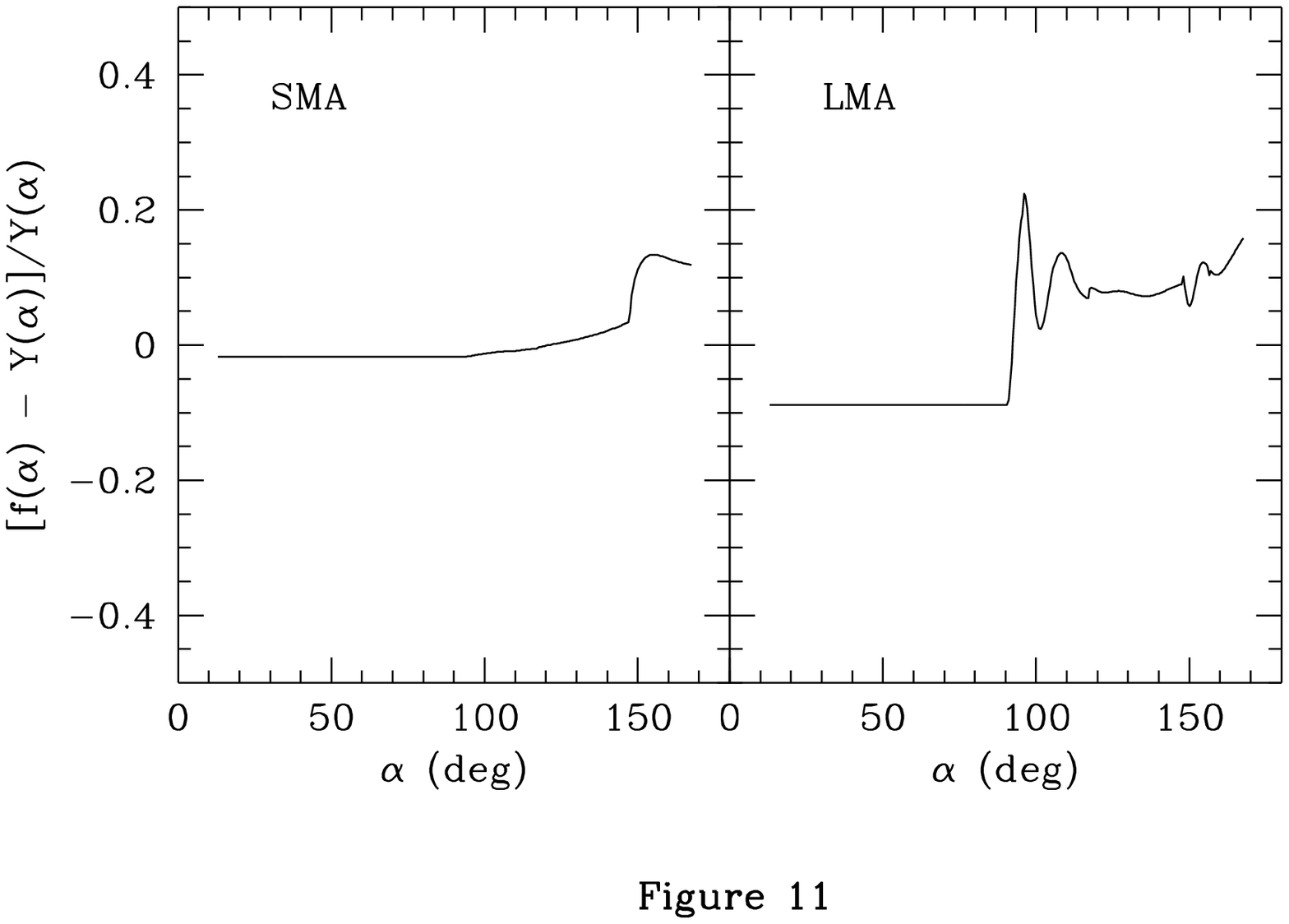,width=5.5in}}
\newpage
\centerline{\psfig{figure=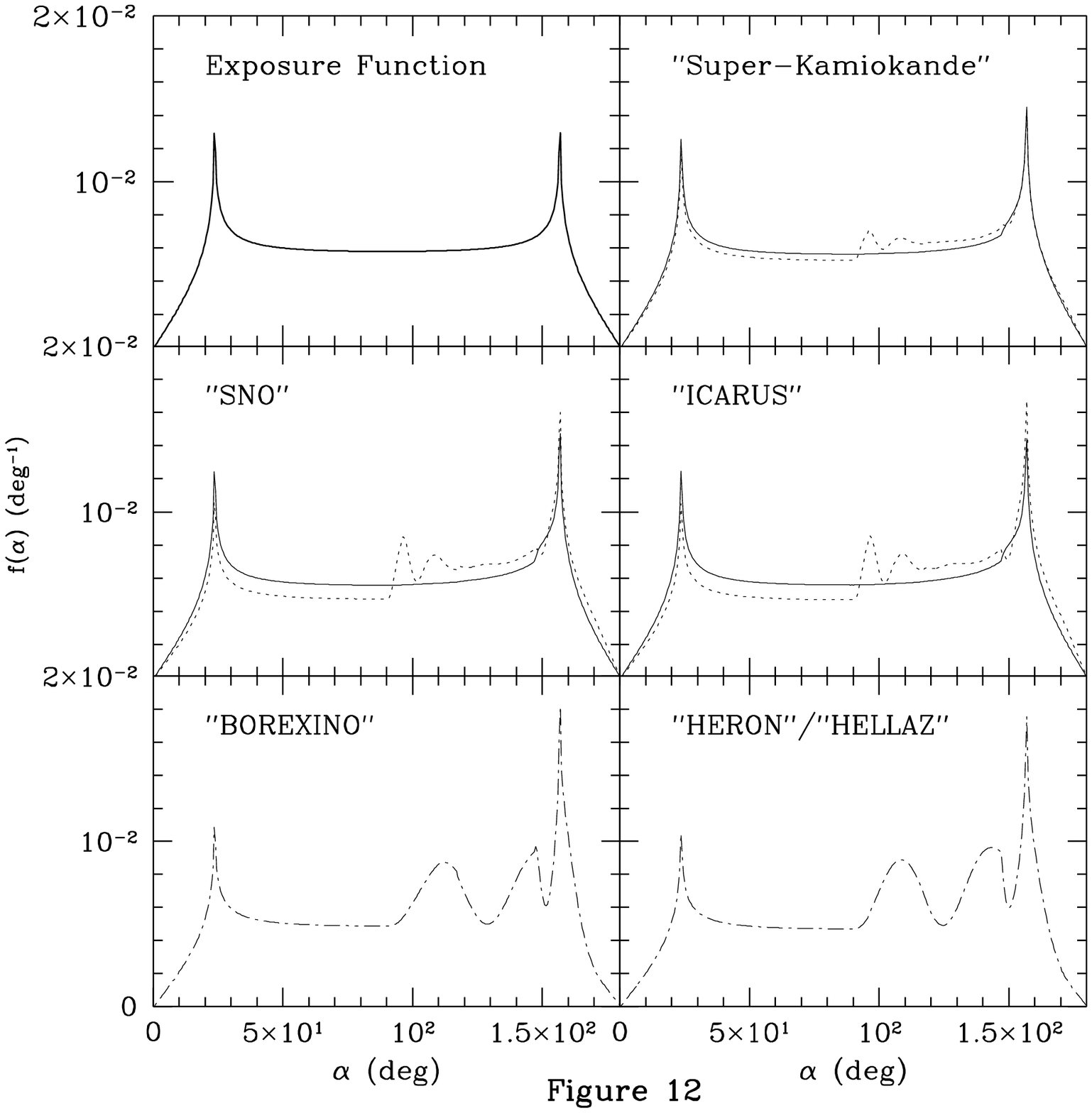,width=5.5in}}
\newpage
\centerline{\psfig{figure=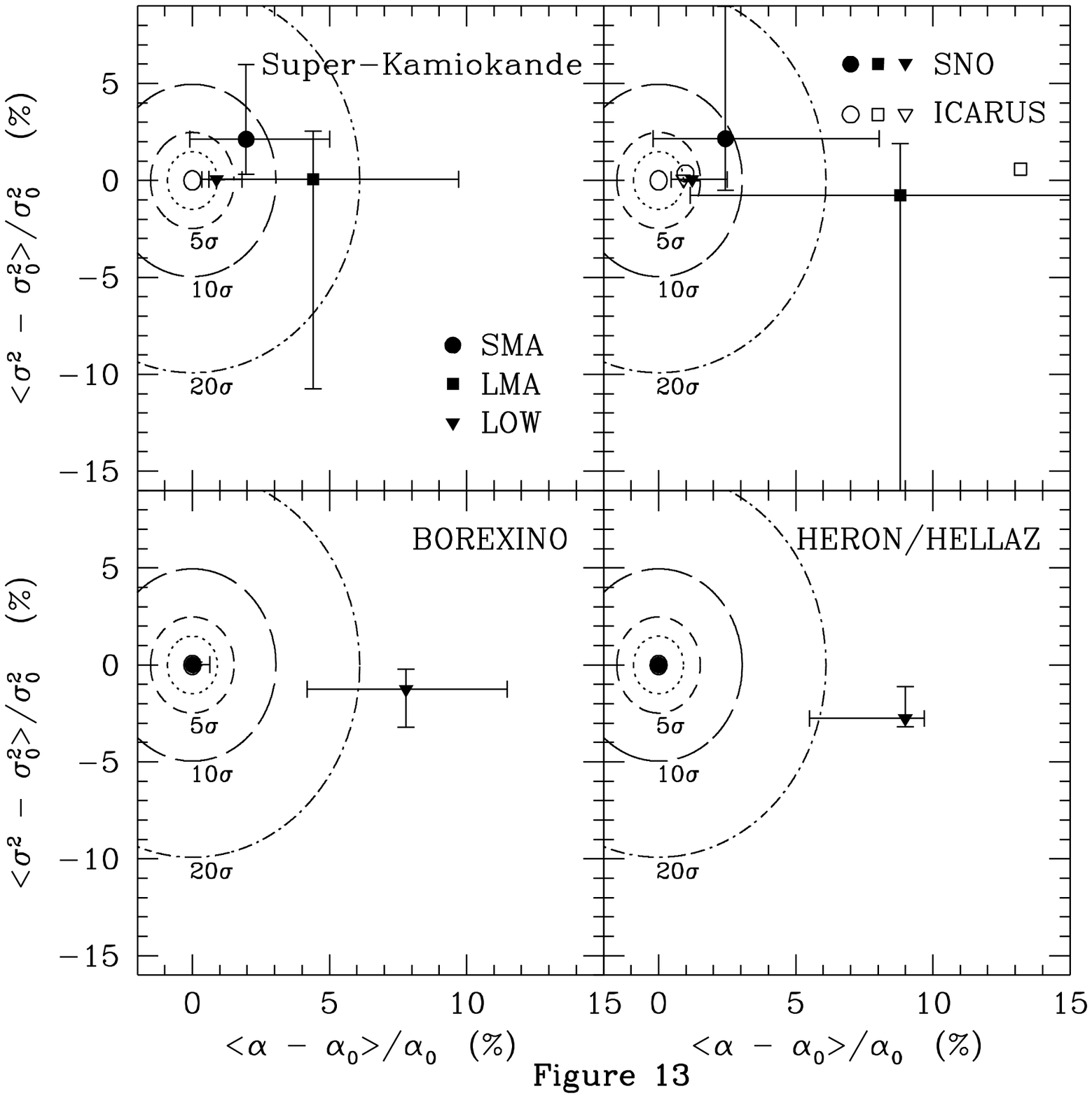,width=5.5in}}
\newpage

\end{document}